\documentclass[10pt,a4paper]{article}

\usepackage{amsfonts,amsmath,amssymb}
\usepackage{amsthm,latexsym}
\usepackage{bbm}
\usepackage[labelformat=simple]{subfig}

\usepackage{bm,multirow}        
\usepackage{graphicx}  
\usepackage{algorithm,algorithmic}

\newtheorem{thm}{Theorem}
\newtheorem{rem}[thm]{Remark}
\newcommand {\av}[1] {\mbox{$\left\{\!\!\left\{ #1 \right\}\!\!\right\}$}}
\newcommand {\jump}[1] {\mbox{$\left[\!\left[ #1 \right]\!\right]$}}

\graphicspath{{./Figures/}}

\usepackage{hyperref}

\begin{document}

\title{Pricing European and American Options Under Heston Model Using Discontinuous Galerkin Finite Elements}

\author{Sinem Kozp{\i}nar\thanks{Institute of Applied Mathematics, Middle East Technical University, Ankara-Turkey
{ksinem@metu.edu.tr}}, Murat Uzunca\thanks{Department of Mathematics, Sinop University, Sinop-Turkey {muzunca@sinop.edu.tr}},
B\"ulent Karas\"ozen\thanks{Institute of Applied Mathematics \& Department of Mathematics, Middle East Technical University, Ankara-Turkey {bulent@metu.edu.tr}}}

\date{}

\maketitle

\begin{abstract}

This paper deals with pricing of European and American options, when the underlying asset price follows Heston model, via the interior penalty discontinuous Galerkin finite element method (dGFEM). The advantages of dGFEM space discretization with Rannacher smoothing as time integrator with nonsmooth initial and boundary conditions are illustrated for European vanilla options, digital call and American put options. The convection dominated Heston model for vanishing volatility is efficiently solved utilizing the adaptive dGFEM. For fast solution of the linear complementary problem of the American options, a projected successive over relaxation (PSOR) method is developed with the norm preconditioned dGFEM. We show the efficiency and accuracy of dGFEM for option pricing by conducting comparison analysis with other methods and numerical experiments.\\

\noindent\textbf{\textit{Keywords:}} Heston model, European option, American option, discontinuous Galerkin method, Rannacher smoothing, preconditioning.

\end{abstract}

\section{Introduction}
\label{sec:intro}

In 1973, Black and Scholes \cite{BlackScholes}
proposed a celebrated model under which the underlying
price follows a geometric Brownian motion with constant volatility, and evaluate European options  via an analytical formula. In spite of its simplicity and mathematical
tractability, the Black-Scholes model is revealed to be unsuccessful, e.g., in predicting the volatility smirk due to the assumption of constant volatility. This important
shortcoming can be overcame by regarding volatility as a source of randomness as in Heston model. Heston model treats the volatility as a square-root process, while  also providing analytical prices for some options such as European vanilla options.

Heston model is  a  two-dimensional reaction-convection-diffusion (RCD) partial differential equation (PDE) with variable coefficients \cite{Heston93,Lipton01mathematical}. The diffusion matrix contains the cross-diffusion term as a result of the correlation between the volatility and the underlying security. The commonly used method for option pricing under the Heston model is the finite difference method (FDM).
It is relatively simple to implement, but the accuracy of the method is limited by
time and space. For the effective numerical solution of these systems,  FDMs with alternating direction implicit (ADI) time-stepping methods
\cite{Hout10adi,Kluge02}, and high-order compact finite difference scheme in space \cite{During17} haven be developed.
When the boundary conditions, the shape of the domain, or data have limited regularity, finite elements methods (FEMs) perform better than FDMs
 \cite{chen14haf,Winkler01}.
Radial basis  \cite{Ballestra13} and spectral methods \cite{Pindza13} are more accurate than the classical FDMs and FEMs, but they require the inversion of full system matrices.
  Fourier-based integration methods (COS) make use of the characteristic function, i.e., the Fourier transform of
the probability density function of the underlying stock
price of Heston model~\cite{Heston93}, which are fast
and accurate \cite{oosterlee11}. Recently wavelet methods are developed using the B-splines basis \cite{Oosterlee13} and Shannon wavelets inverse Fourier methods \cite{Oosterlee16}. Due to the local nature of the wavelet basis, they are more robust and efficient than the COS methods. For American options, the value function satisfies a parabolic partial differential variational inequality system due to the early exercise constraint. FDMs are combined with operator splitting \cite{clarke1999multigrid,Hout15,Ikonen08,ikonen2009operator,oosterlee2003multigrid,Safaei18}  to the  linear complementarity problem  (LCP). FEM is also applied for solving American options under Heston model \cite{Feng11,kunoth2012multiscale,zvan1998penalty}.

This paper deals with the numerical computation of European and American option pricing problems under the Heston model with discontinuous Galerkin finite elements method (dGFEM). The basis functions in dGFEM are discontinuous along the inter-element boundaries in contrast to the classical FEMs. The dGFEM has a number of desirable properties like the weakly enforcement of the boundary conditions, $hp$ (space and order) adaptivity and parallelization. In contrast to the stabilized continuous Galerkin finite elements methods for PDE with convection term like the Heston model, discontinuous Galerkin (dG) methods
produce stable discretization without the need for extra stabilization strategies
and damp the unphysical oscillations for convection dominated problems.  The
dG combines the best properties of the finite volume and continuous finite elements
methods. Finite volume methods can only use lower degree polynomials, and continuous
finite elements methods require higher regularity due to the continuity requirements. The disadvantage of the dG methods is higher condition numbers of the matrices of the associated algebraic systems \cite{Uzunca14} than for finite elements or
finite differences. Several preconditioning techniques are developed as a remedy. One of them is applied here for American options.

We apply the symmetric interior penalty Galerkin (SIPG)  method \cite{DNArnold_FBrezzi_BCockburn_LDMarini_2000a,riviere08dgm} with upwinding for the convective part \cite{Ayuso09}  to Heston model for some European options, such as vanilla and digital call options, and for American options.  The SIPG method is the most used and popular method among the dG methods.
On the other hand, other types of  dGFEMs are applied to option pricing; pricing of European and American options with Constant Elasticity of Variance (CEV) model  \cite{Nicholls15} and American options for the Black-Scholes equation  \cite{Song17},  nonsymmetric interior penalty Galerkin (NIPG) method for pricing of European options under Black-Scholes  \cite{Hozman14}, Heston models \cite{Hozman16}, and Asian options \cite{Hozman17,Hozman16a}.
An interesting feature of the Heston model is the occurrence of sharp layers or discontinuities for vanishing volatility. In the case of European call options, for instance, extremely high foreign interest rates make the problem convection dominated \cite{Kluge02} when the underlying volatility is very small. In these cases, the naive approach is to refine the spatial mesh uniformly, which increases the degrees of freedom and refines the mesh unnecessarily in regions where the solutions are smooth. In some FDMs and FEMs \cite{During14,Hout10adi,Kluge02},  predefined nonuniform grids are used to resolve the discontinuities or the sharp layers accurately. In practice, the location of the interior or boundary layers for convection dominated problems are usually not known a priori. Adaptive methods can detect the layers using a posteriori error estimators  and refine the mesh locally. Due to the local nature of the basis functions of the dGFEMs, the sharp layers and the singularities of the solution can be detected easily using the adaptive techniques \cite{Schotzau09}. We applied adaptive dGFEM based on a posteriori error estimator for an accurate and efficient solution of convection dominated Heston model for  European options.

Option pricing models have nonsmooth initial data, with the discontinuous first derivatives of the payoff functions. The most popular time discretization method in option pricing is the Crank-Nicolson (CN) method, which leads undesired oscillations for nonsmooth initial data. The instability of the CN method is remedied by applying in the first four steps the implicit backward Euler (BE) method and then continuing with CN method as time integrator, known as Rannacher smoothing \cite{Rannacher84}. We apply the Rannacher smoothing for the Heston model with combination of SIPG and CN. The effect of the Rannacher smoothing is clearly visible for digital call options.
 For American options, the LCP is solved by projected successive over relaxation (PSOR) \cite{Ikonen08,ikonen2009operator} or by projected Gauss-Seidel  (PSG) method. At each time step, a large full LCP has to be solved \cite{clarke1999multigrid}.  Because the condition number of the dGFEM discretized matrices increases rapidly for finer meshes, the convergence of the PSOR method slows down. Using the norm preconditioner in \cite{Georgoulis08} designed for dGFEM discretization of RCD equations, we show that the convergence of PSOR method can be accelerated.  Numerical results show that the number of iterations at each time step is essentially mesh independent. To the best of our knowledge, the PSOR with a matrix preconditioner is used for the first time in the evaluation of American options.  We further present the detailed comparison of the SIPG for European and American options with radial basis functions with the partition of unity methods (RBF-PUM) \cite{Mollapourasl19}, and with the NIPG \cite{Hozman16}.

The paper is organized as follows: In the next section, we introduce the Heston model for option pricing and give strong and variational forms of the underlying PDE. The space discretization via SIPG method and time discretization by CN method with Rannacher smoothing are described in Section~\ref{sec:dg}.
The American option pricing as a LCP is formulated in Section~\ref{sec:aolcp} with the preconditioned PSOR method. In Section~\ref{sec:numex}, we give numerical orders of convergence for a test problem with a known solution, and  we report on numerical results for European call as well as digital call and American put options, convection dominated problem for European call options using adaptive dGFEM. At the end of Section~\ref{sec:numex} we compare the dGFEM with the RBF-PUM methods for European and American options. The paper ends with some conclusions.

\section{Heston model}
\label{sec:heston}

Heston stochastic volatility model assumes that the value of the underlying security $S_t$ is governed by the stochastic differential equation~\cite{Heston93}
\begin{equation*}
	dS_t=(r_d-r_f)S_t dt+\sqrt{v_t}S_t dW^{S}_t,
\end{equation*}
for which the variance $v_t\geq 0$ follows the square-root process
\begin{equation*}
	dv_t=\kappa(\theta-v_t)dt+\sigma\sqrt{v_t}dW^{v}_t,
\end{equation*}
where $W^{S}_t$ and $W^{v}_t$ are standard Brownian motions with constant correlation $\rho\in (-1,1).$ Here, $r_d$ is the domestic interest rate, $r_f$ is the foreign interest rate, $\kappa$ is the mean reversion rate, $\theta$ is the long-run mean level of $v_t,$ and $\sigma$ is the volatility of the volatility. We remark that
the variance process $v_t$ becomes strictly positive if the so-called Feller condition
$
2\kappa\theta\geq \sigma^2
$
is satisfied. However, as  experienced in some practical applications,  the variance process $v_t$ can reach zero for some  $t>0$ with probability 1.  But  even this condition is not fulfilled, the Lebesgue measure over the set of times at which $v_t=0$ is zero. Therefore, one can conclude that the variance process will not be stuck at zero even when the Feller condition does not hold (see~\cite{Andersen2007} for a more detailed discussion).

Remarkably, the Heston model provides a Fourier-based pricing formula for European vanilla~\cite{Heston93} and digital options~\cite{Lazar03}. Unfortunately, the Heston model does not yield a similar, easily computable pricing formula for American options.

\subsection{Heston model as parabolic PDE}
\label{HestonPDE}

As a direct application of the Feyman-Kac theorem, the no-arbitrage price of a European option under the Heston model can be characterized by a two-dimensional RCD  equation with variable coefficients. More precisely,  the European option price $U(\tau,v,x)$ with strike $K$ and maturity $T$ satisfies the following linear two-dimensional variable coefficient RCD  equation~\cite{Heston93,Lipton01mathematical}
\begin{equation}\label{pdex1}
	\frac{\partial U}{\partial \tau}- \mathcal{J}_{\tau}^x U+r_dU=0,
\end{equation}
for all  $v>0,$ $x\in(-\infty,\infty)$ and $\tau\in(0,T]$ with the initial condition
\begin{equation*}
U(0,v,x)=U^0(v,x),
\end{equation*}
where $x=\log(S/K),$ $\tau=T-t$ is the time to maturity $T$ at time $t$, and  $U^0(v,x)$ denotes the payoff function. The linear operator $\mathcal{J}_{\tau}^x$ is defined by
\begin{align*}\label{diffconvrecx}
	\mathcal{J}_{\tau}^x U&=\frac{1}{2}v\frac{\partial^2 U}{\partial
		x^2}+(r_d-r_f-\frac{1}{2}v)\frac{\partial U}{\partial x}
	+\rho\sigma v \frac{\partial^2 U}{\partial
		v\partial{x}}
	+\frac{1}{2}\sigma^2 v\frac{\partial^2 U}{\partial
		v^2}+(\kappa(\theta-v)-\lambda(t,v,x))\frac{\partial
		U}{\partial v},
\end{align*}
where $\lambda(\tau,v,x)$ represents the market price of volatility
		risk, which is commonly chosen as $\lambda(\tau,v,x)=\lambda_0 v$  for some constant $\lambda_0$. Note that we can write
$$
\kappa(\theta-v)-\lambda_0 v= (\kappa+\lambda_0)\left(\frac{\kappa}{\kappa+\lambda_0}\theta-v\right).
$$
 By modifying the parameters $\kappa$ and $\theta$ as $\kappa^*=\kappa+\lambda_0$ and $\theta^*=\kappa\theta /(\kappa+\lambda_0)$, the constant $\lambda_0$ can be  eliminated.
Therefore, the market price of volatility risk is commonly set to zero. Throughout the paper, we follow this approach and assume that $\lambda_0=0$.

Different from its European counterparts, American options provide their holders the flexibility of exercising at any time up to maturity. Therefore, when evaluating American options, one should take into account its early exercise feature which addresses a LCP. Suppose that  $U(\tau,v,x)$ denotes the price of an American option with  strike $K$ and maturity $T.$ As is well-known in the literature (see, e.g.~\cite{Hout15}),  the American option price   is defined as the solution of the following LCP
\begin{equation}\label{pdex1_American}
	\begin{aligned}
		&\frac{\partial U}{\partial \tau}-\mathcal{J}_{\tau}^{x}U+r_dU\geq 0, ~U\geq U^0,\\
		&\left(\frac{\partial U}{\partial \tau}-\mathcal{J}_{\tau}^{x}U+r_dU\right)(U-U^0)=0,
	\end{aligned}
\end{equation}
with the payoff function $U^0(v,x)$.

We consider an open bounded domain $\Omega$ with the boundary $\Gamma = \Gamma_D\cup\Gamma_N$, where on $\Gamma_D$ the Dirichlet  and  on $\Gamma_N$  the Neumann boundary conditions are prescribed, respectively. Then, the log-transformed PDE \eqref{pdex1} for European options is expressed as the following RCD equation
\begin{subequations}\label{convdiff_x}
	\begin{align}
		\frac{\partial U}{\partial \tau}-\nabla\cdot (A \nabla U)+ {\mathbf b}\cdot\nabla U+r_dU&=0  & \text{in } (0,T)\times\Omega ,\label{convdiff_xmodel}\\
		U(\tau,\mathbf{z}) &= U^D(\tau,\mathbf{z})  & \text{on } (0,T)\times\Gamma_D, \label{convdiff_xmodelDB}\\
		A\nabla U(\tau,\mathbf{z})\cdot\mathbf{n} &= U^N(\tau,\mathbf{z})  & \text{on } (0,T)\times\Gamma_N,\label{convdiff_xmodelNB}\\
		U(0,\mathbf{z}) &= U^0(\mathbf{z})  & \text{in } \{0\}\times\Omega ,\label{convdiff_xmodel_init}
	\end{align}
\end{subequations}
where $\mathbf{n}$ is the outward unit normal vector, $\mathbf{z}=(v,x)^T$,  throughout this paper, is the spatial element.
In \eqref{convdiff_x}, the diffusion matrix and convection field  are given by
\begin{equation*}
	A  =
	\frac{1}{2}v\left( \begin{array}{ccc}
		\sigma^2  & \rho\sigma \\
		\rho\sigma  & 1
	\end{array} \right)\quad\text{and} \quad {\mathbf b} =
	v \left( \begin{array}{c}
		\kappa\\
		\frac{1}{2}
	\end{array}\right) +
	\left( \begin{array}{c}
		-\kappa\theta+\frac{1}{2}\sigma^2\\
		-(r_d-r_f)+ \frac{1}{2}\rho\sigma
	\end{array}\right).
\end{equation*}

\begin{rem}
	Although the transformed PDE \eqref{pdex1} is defined on the computational domain $(0,\infty)\times(-\infty,\infty)$, dGFEM  must be performed  on a bounded spatial region $\Omega=(v_{\text{min}},v_{\text{max}})\times(x_{\text{min}},x_{\text{max}})$ for the numerical simulations.
	The spatial domain is truncated in practice based on the standard financial arguments, such that the
	error caused by truncating the solution domain has a negligible effect on the option
	values in the region of interest.
\end{rem}

\subsection{Variational form of European options}
\label{variationalform}

Let $L^2(\Omega)$ be the space consisting of all square integrable functions on $\Omega$, and $H^1(\Omega)$ be the Hilbert space of those functions in $L^2(\Omega)$  having square integrable first-order partial derivatives, with the subspace $H_0^1(\Omega )$ including functions with zero trace on the boundary. The variational (weak) form of the PDE \eqref{convdiff_x} of European options is obtained by multiplying it with a test function $w\in H_0^1(\Omega )$ and integrating by parts over the domain $\Omega$. Then, for a.e. $\tau\in(0,T]$, we seek a solution $U(\tau,v,x)\in H_D^1(\Omega ):=\{U\in H^1(\Omega): U=U^D\:\: \text{on} \:\:\Gamma_D\}$ satisfying

\begin{subequations}\label{cont_weak}
	\begin{align}
		\int_{\Omega} \frac{\partial U}{\partial \tau}w d\mathbf{z} + a(U,w) &= \int_{\Gamma_N} U^Nw ds,  &\forall w\in H_0^1(\Omega ),\\
		\int_{\Omega} U(0,\mathbf{z})wd\mathbf{z} &=  \int_{\Omega}U^0 wd\mathbf{z},  &\forall w\in H_0^1(\Omega ),
	\end{align}
\end{subequations}
where $ds$ is the arc-length element on the boundary. In \eqref{cont_weak}, $a(U,w)$ is the classical bilinear form given by
$$
a(U,w)=\int_{\Omega}\left( A \nabla U\cdot \nabla w + {\mathbf b }\cdot\nabla Uw+r_dUw\right)d\mathbf{z}, \quad \forall w \in H_0^1(\Omega).
$$
For the existence and  uniqueness of a solution, we assume that the matrix $A$ is positive definite, i.e. $v >0$ and $\rho \in (-1,1)$ which is usually satisfied. Then there exist constants $C$,  $c_1$ and $c_2$ for all $U$ and $w$, so that the bilinear  form $a(\cdot,\cdot)$ is continuous and weakly coercive \cite{Winkler01}
\begin{align}
	|a(U,w)| & \le C ||U||_{H^1(\Omega )}||w||_{H^1(\Omega )}, &U, w \in H^1(\Omega ),  \label{bil_cont}\\
	a(U,U) &\ge c_1 ||U||_{H^1(\Omega )}^2 - c_2 ||U||^2_{L^2(\Omega )}, &U \in H^1(\Omega ). \label{bil_gard}
\end{align}
The inequality \eqref{bil_cont} accounts to the continuity of the bilinear form and the inequality \eqref{bil_gard} is the G{\aa}rding inequality. The weakly coercive bilinear form $a(\cdot,\cdot)$ can be transformed into a coercive one using the substitution $\tilde{U} = e^{c_2\tau}U$ \cite[Sec.~1.3.3]{solin2005partial}.
Then, there exist a unique solution to the problem \eqref{cont_weak} and the following energy estimate holds \cite{Burkovska16,Winkler01}
$$
\max_{\tau \in[0,T]} ||U(\tau,\mathbf{z})||_{L^2(\Omega)}^2 +  c_1 \int_0^T ||U(\tau,\mathbf{z})||_{H^1(\Omega)}^2 \le ||U(0,\mathbf{z})||_{L^2(\Omega)}^2.
$$
The existence of a solution
strongly depends on which space the solution sought to be in.
For the existence of a solution, the integrability of the solution is required, i.e. their norms must be finite as above if $v >0$.  When  $v\approx 0$, the diffusion matrix $A$ tends to zero matrix and spatial derivative of $U$ can become arbitrarily large without violating the integrability. 
It was shown in \cite{Kufner87} for the existence and uniqueness of the solution with singularities, the theory of (regular) Sobolev spaces can be conveyed to weighted spaces.

\section{Symmetric interior penalty Galerkin method}
\label{sec:dg}

The interior penalty Galerkin (IPG) methods are well-known members of the family of dG methods, which use discontinuous polynomial approximations and enforce boundary conditions weakly \cite{riviere08dgm}. There are seldom works using dG methods in option pricing. In \cite{Hozman16}, a nonsymmetric variant of IPG method was used to price the European option under one dimensional Black-Scholes as
a PDE, and the same method was extended to two-dimensional PDE case for the valuation of Asian options in \cite{Hozman16a}. In both studies, upwinding is used for convective terms. In this paper, we use SIPG method with upwinding for the convective term, to price European and American options under the Heston model.

Let $\{\xi_{h}\}_h$ be a family of disjoint partition of the domain $\Omega$ into shape regular (triangular) elements $K$, i.e. $\Omega = \cup_{K\in \xi_{h}}K$. We set the mesh-dependent finite dimensional solution and test function space by
$$
W_{h}:=W_{h}(\xi_{h})=\left\{ w\in L^2(\Omega ) : w|_{K}\in\mathbb{P}_k(K) ,\; \forall K\in \xi_{h} \right\},
$$
where $\mathbb{P}_k(K)$ denotes the space of all polynomials up to degree $k$ defined on the element $K$. The functions in $W_{h}$ are discontinuous along the inter-element boundaries, which leads to the fact that on an interior edge $e$ shared by two neighboring triangles $K$ and $K'$ in $\xi_{h}$, there are two different traces from either triangles. Thus, we define the jump and average operators of a function $w\in W_{h}$ on $e$, respectively, by
$$
\jump{w}_e= w|_{K}\mathbf{n}_{K} + w|_{K'}\mathbf{n}_{K'}\; , \quad \av{w}_e=\frac{1}{2}(w|_{K} + w|_{K'}),
$$
where $\mathbf{n}_K$ denotes the exterior unit vector on the boundary of $K$. On a boundary edge $e\subset\partial\Omega$, we set $\jump{u}_e= u|_{K}\mathbf{n}$ and $\av{u}_e=u|_{K}$. In addition, we form the sets of inflow and outflow edges as the following
$$
\Gamma^- = \{ \mathbf{z}\in\partial\Omega : \mathbf{b}(v)\cdot\mathbf{n}(v,x)<0\} \; , \qquad \Gamma^+ = \partial\Omega\setminus\Gamma^- ,
$$
$$
\partial K^- = \{ \mathbf{z}\in\partial K: \mathbf{b}(v)\cdot\mathbf{n}_K(v,x)<0\} \; , \qquad \partial K^+ = \partial K\setminus\partial K^-.
$$
Moreover, we denote by $\Gamma_{h}^0$, $\Gamma_{h}^D$ and $\Gamma_{h}^N$ the sets of interior, Dirichlet boundary and Neumann boundary edges, respectively, and we set $\Gamma_{h}=\Gamma_{h}^0\cup\Gamma_{h}^D$. Then, in space IPG discretized semi-discrete system of the PDE \eqref{convdiff_x} reads as: for a.e. $\tau\in (0,T]$, for all $w_h\in W_{h}$, find $U_h:=U_h(\tau,\mathbf{z})\in W_{h}$ such that
\begin{subequations} \label{dg}
	\begin{align}
		\int_{\Omega}\frac{\partial U_{h}}{\partial \tau}w_{h}d\mathbf{z} + a_{h}(U_{h},w_{h}) &=f_{h}(w_{h}) ,\\
		\int_{\Omega} U_{h}(0,\mathbf{z})w_hd\mathbf{z} &= \int_{\Omega} U^0w_hd\mathbf{z},
	\end{align}
\end{subequations}
with the bilinear and linear forms given by
\begin{align*}
		a_{h}(U_{h}, w_{h})=& \sum \limits_{K \in {\xi_{h}}} \int_{K} \left( A \nabla U_{h}\cdot\nabla w_{h} + {\mathbf b} \cdot \nabla U_{h} w_{h} + r_dU_hw_h\right)d\mathbf{z} \\
		& + \sum \limits_{ e \in \Gamma_{h}} \int_{e} \left( \frac{\gamma_e}{h_{e}}\jump{U_{h}}_e\cdot\jump{w_{h}}_e - \av{A  \nabla U_{h}}_e \jump{w_{h}}_e + \mu\av{A \nabla w_{h}}_e\jump{U_{h}}_e \right)ds  \\
		& + \sum \limits_{K \in {\xi_{h}}} \left( \int\limits_{\partial K^-\setminus\partial\Omega } {\mathbf b \/} \cdot \mathbf{n}_K (U_{h}^{out}-U_{h})  w_{h} ds -  \int\limits_{\partial K^-\cap \Gamma^{-}} {\mathbf b}\cdot \mathbf{n}_K U_{h} w_{h} ds \right), \\
		f_{h}( w_{h})=&  \sum \limits_{e \in {\Gamma_{h}^N}} \int_{e} U^N w_{h} d\mathbf{z} + \sum \limits_{e \in {\Gamma_{h}^D}} \int_{e} U^D\left( \frac{\gamma_e}{h_{e}}w_{h} + \mu A \nabla w_{h} \right)ds\\
		&  - \sum \limits_{K \in {\xi_{h}}}\int\limits_{\partial K^-\cap \Gamma^{-}} {\mathbf b}\cdot \mathbf{n}_K U^D w_{h} ds,
\end{align*}
where $U_{h}^{out}$ denotes the trace of $U_h$ on an edge $e$ from outside the triangle $K$.
For a positive number $\epsilon$ and the so-called penalty parameter $\gamma$, the parameter $\gamma_e := \epsilon \gamma$ is used for the penalization of the solutions on the edges to ensure the coercivity of the bilinear form, as a result, the positive definiteness of the dG stiffness matrix. In addition, the parameter $\mu$ is the IPG parameter determining the type of the dG scheme: $\mu =1$ corresponds to the NIPG method, $\mu =-1$ to the SIPG method and $\mu =0$ for incomplete interior penalty Galerkin (IIPG) method. In \cite{Hozman16a}, the authors use NIPG with the setting $\mu =1$, and they take $\gamma =1$ and $\epsilon = \sigma $. However, it is known that the NIPG method produces sub-optimal solutions when even higher degree polynomials are used \cite{riviere08dgm}. On the other hand, the SIPG method provides optimal convergence rates for any degree polynomials. In this study, we use the SIPG method to discretize the PDE problem in space. In the SIPG scheme, the penalizing term
should be selected sufficiently large to ensure the coercivity of the bilinear form \cite[Sec. 7.6]{riviere08dgm}. At the same time, it should not be too large since the stiffness matrix becomes ill-conditioned for large penalty parameters. Here, we set the positive parameter $\epsilon$ locally given by
$$
\epsilon_K := \left\|\left|\sqrt{A}\right|_2^2\right\|_{L^{\infty}(K)}, \quad \forall K\in\xi_h,
$$
where $|\cdot|_2$ denotes the matrix $2$-norm. Moreover,
we follow \cite{Epshteyn07} to estimate the penalty parameter $\gamma$, namely above a  threshold value the bilinear form is coercive and the scheme is stable and convergent. From the uniform ellipticity (positive definiteness) of the diffusion matrix $A(v)$, it follows that there exist two constants $d_0$ and $d_1$ such that the following inequality holds
$$
d_0 \bm{y}^T\bm{y} \le \bm{y}^T A \bm{y} \le d_1 \bm{y}^T\bm{y}, \qquad \forall \bm{y}\in\mathbb{R}^2.
$$
Then, one can compute and set the penalty parameter $\gamma$ as  \cite{Epshteyn07}
\begin{align*}
	\gamma  &=  \frac{3d_1^2}{d_0} k(k+1) \cot \theta, & \text{on each interior edge},\\
	\gamma  &=  \frac{6d_1^2}{d_0} k(k+1) \cot \theta, & \text{on each boundary edge},
\end{align*}
where $\theta$ denotes the smallest angle over all triangular elements $K$ in $\xi_{h}$.

The semi-discrete solution of the SIPG discretized system \eqref{dg} of the Heston model is given by
\begin{equation}\label{4}
	U_h(\tau,\mathbf{z})=\sum^{N_e}_{m=1}\sum^{N_k}_{j=1}u^{m}_{j}(\tau) \varphi^{m}_{j}(\mathbf{z}),
\end{equation}
where $\varphi^{m}_{j}$ and $u^{m}_{j}$, $j=1, \ldots, N_k$, $m=1, \ldots, N_e$, are the basis functions spanning the space $W_{h}$ and the unknown coefficients, respectively. The number $N_k$ denotes the local dimension of each dG element with the identity $N_k=(k+1)(k+2)/2$, and $N_e$ is the number of dG elements (triangles), leading to the dG degrees of freedom $N:=N_e\times N_k$. Substituting \eqref{4} into \eqref{dg} and choosing $w_h=\varphi^{m}_{j}, \: j=1, \ldots, N_k$, $m=1, \ldots, N_e$, we obtain the following system of ordinary differential equations (ODEs) for the unknown coefficient vector $\bm{u}:=\bm{u}(\tau) =(u_{1}^{1}(\tau), \ldots , u_{N_k}^{1}(\tau),\ldots , u_{1}^{N_e}(\tau), \ldots , u_{N_k}^{N_e}(\tau))^T\in\mathbb{R}^N$
\begin{equation*}
	{\mathbf M} \bm{u}_{\tau} + {\mathbf A} \bm{u}   = \bm{f},
\end{equation*}
where ${\mathbf M}$ is the mass matrix,  ${\mathbf A}$ is the stiffness matrix and $\bm{f}:=\bm{f}(\tau)$ is the right hand side vector, with the entries $({\mathbf M})_{ij}=(\varphi^{j}, \varphi^{i})_{\Omega}$, $({\mathbf A})_{ij}=a_h(\varphi^{j}, \varphi^{i})$ and $(\bm{f})_{i}=f_h(\varphi^{i})$, $1\leq i,j \leq N$.

For the time discretization, we consider a uniform partition of $[0,T]$ into $J$ time intervals $I_n=(\tau^{n-1},\tau^{n}]$ of length $\Delta \tau$, $n=1,2,\ldots , J$, and we set discrete times $\tau^n=n\Delta \tau$, $n=0,1,\ldots , J$.  We denote by $\bm{u}^n=\bm{u}(\tau^n)$ and $\bm{f}^n=\bm{f}(\tau^n)$ the full discrete solution vector and the right hand side vector at time $\tau^n$, respectively. The presence of discontinuities in the initial conditions creates challenge for the solution of option pricing problems.
The discontinuities pollute the solutions resulting in the reduction of the theoretical rate of convergence of a numerical scheme. When central finite difference discretization is used for the time derivative, like the  CN scheme, spurious oscillations are introduced.  Spurious solutions are removed using the so-called Rannacher time-stepping \cite{Rannacher84}.
It  involves
applying four steps BE method with the step size $\Delta \tau/2$ and continuing with the CN method with the step size $\Delta \tau$. Thus the  same coefficient matrix is formed \cite{Ikonen08}
\begin{equation}\label{fullydisc}
	\begin{aligned}
		\left ({\mathbf M}  + \frac{\Delta \tau }{2}{\mathbf A} \right)\bm{u}^{\frac{m+1}{2}}   &=  {\mathbf M} \bm{u}^{\frac{m}{2}} + \frac{\Delta \tau}{2} \bm{f}^{\frac{m+1}{2}}, &m =0,1,2,3, \\
		\left ({\mathbf M}  + \frac{\Delta \tau }{2}{\mathbf A}  \right)\bm{u}^{n+1} &=  \left ( {\mathbf M}  - \frac{\Delta \tau}{2}  {\mathbf A} \right) \bm{u}^n + \frac{\Delta \tau}{2} (\bm{f}^n +\bm{f}^{n+1}),
		&n =2,3, \ldots
	\end{aligned}
\end{equation}
The coefficient matrix is factorized by LU decomposition at the initial time step and used in all successive time steps, which makes the time integration efficient. A few implicit steps filter out high frequency error components, the loss of accuracy is negligible and overall second-order convergence is retained \cite{Giles06,Rannacher84}. Besides Rannacher smoothing there exist other techniques for handling the discontinuities in the initial data of option pricing models and to increase the convergence rate. A frequently used smoothing technique is constructed by calculating the average
value on the intervals over the function \cite{Kreiss70}. It was used in \cite{Heston00} for vanilla put options.  A fourth-order implementation based on compact FDMs in two dimensions is constructed using inverse Fourier transform \cite{During15}. Another technique is shifting the grid so that the discontinuities are placed in between the grid points \cite{Tavella00}. In this way, the order of convergence can be increased. These methods and Rannacher smoothing are discussed for one and two-factor option pricing problems in \cite{Pooley03}.

\section{American option: linear complementary problem }
\label{sec:aolcp}

For the American put option under Heston model, we consider the LCP \eqref{pdex1_American} in the form
\begin{subequations}\label{pdeA_American}
	\begin{align}
		\frac{\partial U}{\partial \tau} + L^HU &\geq 0, \label{pdeA_American1}\\
		\left(\frac{\partial U}{\partial \tau} + L^HU\right)(U-U^0) &= 0,\label{pdeA_American2}\\
		U &\geq U^0,\label{pdeA_American3}
	\end{align}
\end{subequations}
where the differential operator is $L^HU:= - \nabla\cdot A\nabla U + \mathbf{b} \cdot\nabla U + r_dU$. Let us set the space $W_A:=\{w\in H^1_D(\Omega) \; : \; w\geq U^0 \}$. Then, for any $w\in W_A$, multiplying \eqref{pdeA_American1} by $w-U^0$ (which does not change the inequality sign), taking integral over the domain $\Omega$ and subtracting the integral of \eqref{pdeA_American2} from it, we obtain the variational formulation of \eqref{pdeA_American} as \cite{Hilber13}
\begin{subequations}\label{weak_American}
	\begin{align}
		\int_{\Omega} \left(\frac{\partial U}{\partial \tau} + L^HU \right)(w-U)d\mathbf{z} & \geq 0, & \forall w\in W_A,\\
		U(0,\mathbf{z}) &= U^0. &
	\end{align}
\end{subequations}
When the term $(L^HU,w)_{L^2(\Omega)}$ for any $w\in W_A$ satisfies the  continuity condition \eqref{bil_cont} and  the G{\aa}rding inequality \eqref{bil_gard}, the problem \eqref{weak_American} admits a unique solution. For details of the proof, we refer to (Section 4.3, \cite{Burkovska16}).

Using the notations and formulations for the European option in the previous section, the numerical pricing of American option after SIPG discretization with the Rannacher smoothing leads in matrix-vector form to the following LCPs
\begin{equation}\label{lcp_American}
\begin{aligned}
	{\mathbf B } \bm{u}^{\frac{m+1}{2}} &\ge {\mathbf F}^m, \qquad\quad {\mathbf B } \bm{u}^{n+1} \ge {\mathbf F}^n, \\
	\bm{u}^{\frac{m+1}{2}} &\ge  \bm{u}^0, \qquad \qquad \bm{u}^{n+1} \ge  \bm{u}^0, \\
	(\bm{u}^{\frac{m+1}{2}}-\bm{u}^0)^T ({\mathbf B} \bm{u}^{\frac{m+1}{2}} - {\mathbf F}^m) &= 0, \qquad (\bm{u}^{n+1}-\bm{u}^0)^T ({\mathbf B} \bm{u}^{n+1} - {\mathbf F}^n)= 0 ,\\
	m=0,1,2,3, & \qquad\qquad\quad n=2,3,\ldots , J-1,
\end{aligned}
\end{equation}
where we set
\begin{align*}\label{rannao}
	{\mathbf B} &=  {\mathbf M} + \frac{\Delta \tau }{2} {\mathbf A}, &  \\
	{\mathbf F}^m &=  {\mathbf M}\bm{u}^{\frac{m+1}{2}} + \frac{\Delta \tau }{2}{\bm{f}}^{\frac{m+1}{2}}, &  m=0,1,2,3, \\
	{\mathbf F}^{n} &=  \left ({\mathbf M}-\frac{\Delta \tau}{2}{\mathbf A}\right ) \bm{u}^n + \frac{\Delta \tau }{2}(\bm{f}^n +
\bm{f}^{n+1}), & n=2,3,\ldots, J-1.
\end{align*}

One of the most popular methods for the solution of the LCPs in numerical pricing of American options is the PSOR method. The PSOR method converges for symmetric positive definite matrices ${\mathbf B}$ \cite{Cryer71}.  The PSOR works also for  nonsymmetric but diagonally dominant matrices \cite{Hilber13}. But the matrix  ${\mathbf B }$ in \eqref{lcp_American} resulting from the SIPG discretization is not diagonally dominant.   Additionally, the condition number of the stiffness matrix  ${\mathbf A}$ is of order ${\mathcal O}(h^2)^{-1}$ for SIPG, which leads to slow  convergence of the PSOR. Here we apply the PSOR with the norm preconditioner  ${\mathbf B_s} = \frac{1}{2}({\mathbf B } +  {\mathbf B }^T)$,  which is designed for the dGFEM discretization of RCD problems  \cite{Georgoulis08}. A linear system ${\mathbf B} \bm{u} = \bm{d}$ can be solved using three different preconditioners
\begin{itemize}
	\item left preconditioner :  ${\mathbf B_s}^{-1}{\mathbf B } \bm{u} ={\mathbf B_s}^{-1} \bm{d}$,
	\item right preconditioner :  ${\mathbf B }{\mathbf B_s}^{-1} \bm{v} = \bm{d},\; \bm{v}= {\mathbf B_s}
	\bm{u}$,
	\item split (two-sided) preconditioner : ${\mathbf B_s}^{-1/2}{\mathbf B } {\mathbf B_s}^{-1/2}  \bm{v} ={\mathbf B_s}^{-1/2} \bm{d}, \; \bm{v} = {\mathbf B_s}^{1/2} \bm{u}$
\end{itemize}

The three preconditioned matrices have the same eigenvalues and are  well conditioned. They are also diagonally dominant, and we are able to use the PSOR method. Among the  three preconditioners, the two-sided preconditioner transforms the nonnormal dGFEM discretized matrices of the
RCD equations to normal matrices ${\mathbf B_s}^{-1/2}{\mathbf B } {\mathbf B_s}^{-1/2}= {\mathbf I} + {\mathbf S}$, where ${\mathbf I}$ is the identity matrix and the matrix ${\mathbf S}$ is skew-symmetric \cite{Georgoulis08}. The preconditioned PSOR algorithm at an individual time step for the LCP \eqref{lcp_American} with the two-sided norm preconditioner ${\mathbf B_s}^{-1/2}{\mathbf B } {\mathbf B_s}^{-1/2}$ is given in Algorithm~\ref{psor}.

\begin{algorithm}[htb!]
	\caption{The two-sided preconditioned PSOR algorithm\label{psor}}
	{\bf Aim:} Finding the unknown solution $\bm{u}^{n+1}$ of the LCP \eqref{lcp_American}.\\
	{\bf Input:} Solution $\bm{u}^{n}$, matrix $\tilde{{\mathbf B}}:= {\mathbf B_s}^{-1/2}{\mathbf B } {\mathbf B_s}^{-1/2}$, vector $\bm{d}:=\mathbf{B}_s^{-1/2}\mathbf{F}^n$.\\[0.3cm]
	Choose an initial guess $\bm{y}^{(0)}\ge {\mathbf B_s}^{1/2}\bm{u}^0$ (possibly $\bm{y}^{(0)}:={\mathbf B_s}^{1/2}\bm{u}^{n}$)\\
	Choose a tolerance $\varepsilon > 0$, and $\omega \in (0,1)$ according to \eqref{omega}.\\
	\begin{algorithmic}
		\FOR{$l=1,2, \ldots$}
		\FOR{$i=1,2, \ldots , N$}
		\STATE $\tilde{\bm{y}} = \frac{1}{{\tilde{\mathbf{B}}}_{ii}}\left ( \bm{d}_i - \sum_{j=1}^{i-1}  {\tilde{\mathbf{B}}}_{ij}\bm{y}_j^{(l)} - \sum_{j=i+1}^{N}  {\tilde{\mathbf{B}}}_{ij}\bm{y}_j^{(l-1)}\right )$
		\STATE $\bm{c}_i = \bm{y}_i^{(l-1)} + \omega(\tilde{\bm{y}} -\bm{y}_i^{(l-1)})$
		\STATE $\bm{y}_i^{(l)} =\max \{({\mathbf B_s}^{1/2}\bm{u}^0)_i,\bm{c}_i \}$
		\ENDFOR
		\IF {$\|\bm{y}^{(l)}-\bm{y}^{(l-1)}\|<\varepsilon$}
		\STATE {\bf stop}
		\ENDIF
		\ENDFOR
		\STATE{$\bm{u}^{n+1}={\mathbf B_s}^{-1/2}\bm{y}^{(l)}$}
	\end{algorithmic}
\end{algorithm}

The optimal over-relaxation parameter $\omega$, which is crucial for the convergence, in the PSOR is chosen according to \cite{ikonen2009operator} as
\begin{equation}\label{omega}
	\omega = \frac{2}{1 + \sqrt{1-\rho_{\mathbf G}^2}}
\end{equation}
where $\rho_{\mathbf G}$ is the spectral radius of the Jacobi iteration matrix ${\mathbf G}= {\mathbf D}^{-1}({\tilde{\mathbf B}}-{\mathbf D})$ with ${\mathbf D}$ as the diagonal of ${\tilde{\mathbf B}}={\mathbf B_s}^{-1/2}{\mathbf B } {\mathbf B_s}^{-1/2}$.

\section{Numerical results}
\label{sec:numex}

In this section, we present numerical results for European and American options to show the accuracy and efficiency of our numerical schemes. All simulations are performed on a Windows 10 machine with a processor Intel Core i7, 2.5 GHz and 8 GB RAM using MATLAB R2014.

In the first problem, we test the numerical convergence orders of our schemes for Heston model with a given true solution. As the second test example, we consider European call options for which semi-analytical
solutions can be obtained. The third test example is the convection dominated European call option pricing model solved by the adaptive dGFEM. It is widely known that the Heston model can be viewed as a convection dominated PDE for low volatility; the numerical solutions exhibit oscillations around $v\approx 0.$  In the case of European call options, the Heston model is convection dominated especially for high foreign interest rates.
Next, we show the performance of the SIPG discretization with Rannacher smoothing in time for digital options with more non-smooth initial data than the European option. Moreover, we solve numerical examples with American put options with the preconditioned  PSOR method, and finally, we compare dGFEM with the RBF method \cite{Mollapourasl19}.

There exists no a consensus in the literature regarding the boundary conditions~\cite{zhu2011predictor}.   From the computational standpoint, when the parameters $\kappa,$ $\sigma$ and $\theta$ of the squared-root process $v_t$ are not chosen according to the Feller condition $2\kappa\theta\geq\sigma^2,$  $v_t$ can become zero for some points in time. In many market situations Feller condition is violated and  boundary condition is imposed at $v=0$ (see for instance~\cite{chen14haf,Ikonen08,Winkler01}). Thereby,  an appropriate boundary condition at $v=0$ is required to solve RCD equation  \eqref{convdiff_x} and LCP \eqref{pdeA_American}.
On the other hand, from the financial standpoint, the Feller condition is satisfied; therefore, only boundary conditions at $x_{\text{min}},$ $x_{\text{max}}$ and $v_{\text{max}}$ are needed. In this case, the RCD equation \eqref{convdiff_x}
and the LCP \eqref{pdeA_American} is subject to an outflow boundary at $v=0,$ see e.g. \cite{Ballestra13,Heston93,Hout10adi}.  Also other types of boundary conditions can be applied. Therefore, it is not certain at which boundary condition Heston PDE and LCP  should be used to achieve highly accurate solutions.
For a  detailed discussion on boundary conditions in option pricing, we refer to~\cite{zhu2011predictor}.

\subsection{Convergence rates of dGFEM approximation}
\label{testex1}

The convergence orders of the SIPG method with BE and  CN time integrators are investigated for the Heston model with a known solution \cite{chen14haf}. The exact smooth solution satisfying the non-homogeneous Dirichlet boundary conditions is given by \cite{chen14haf}
$$
U(\tau ,v,x) = e^{-\tau}\cos (\pi v) \cos (\pi x) ,
$$
in the domain $\Omega=[0,4]\times [-2,2].$ The parameters are listed in Table~\ref{Dirichlet_parameters} \cite{chen14haf}.
\begin{table}[htb!]
	\centering
	\caption{Parameter set for the Example~\ref{testex1}}
	\label{Dirichlet_parameters}
	\begin{tabular}{l*{7}{c}r}
		\hline\noalign{\smallskip}
		$\kappa$ & $\theta$ & $\sigma$ & $\rho$ & $r_d$ & $r_f$  & $T$\\
		\noalign{\smallskip}\hline\noalign{\smallskip}
		1.0 & 0.09 & 0.4 & -0.7 & 0.05 & 0.01 &  1  \\
		\noalign{\smallskip}\hline
	\end{tabular}
\end{table}

The spatial step size and temporal step size are denoted by $h = \Delta v = \Delta y$ and $\Delta \tau$, respectively.
We present the approximation errors and numerical convergence rates for the SIPG discretization with linear and quadratic polynomials, and with the time integrators  BE and CN. In order to measure the error $U(\tau , {\mathbf z} ) - U_h(\tau , {\mathbf z})$ between the true solution and the approximate solution, we use the following $L^2(L^2)$ ($L^2$-norm in both space and time) and  $L^2(H^1)$ ($L^2$-norm in time, $H^1$ semi-norm in space) type norms
\begin{align*}
\|w( \tau ,{\mathbf z} )\|_{L^2(L^2)} &= \left( \Delta\tau \sum_{n=1}^J \|w( \tau^n,{\mathbf z} )\|^2_{L^2(\Omega)} \right)^{1/2}, \\
\|w( \tau ,{\mathbf z} )\|_{L^2(H^1)} &= \left( \Delta\tau \sum_{n=1}^J \| \nabla w( \tau^n,{\mathbf z} )\|^2_{L^2(\Omega)} \right)^{1/2},
\end{align*}

\begin{figure}[htb!]
	\centering
	\subfloat{\includegraphics[width=0.53\linewidth]{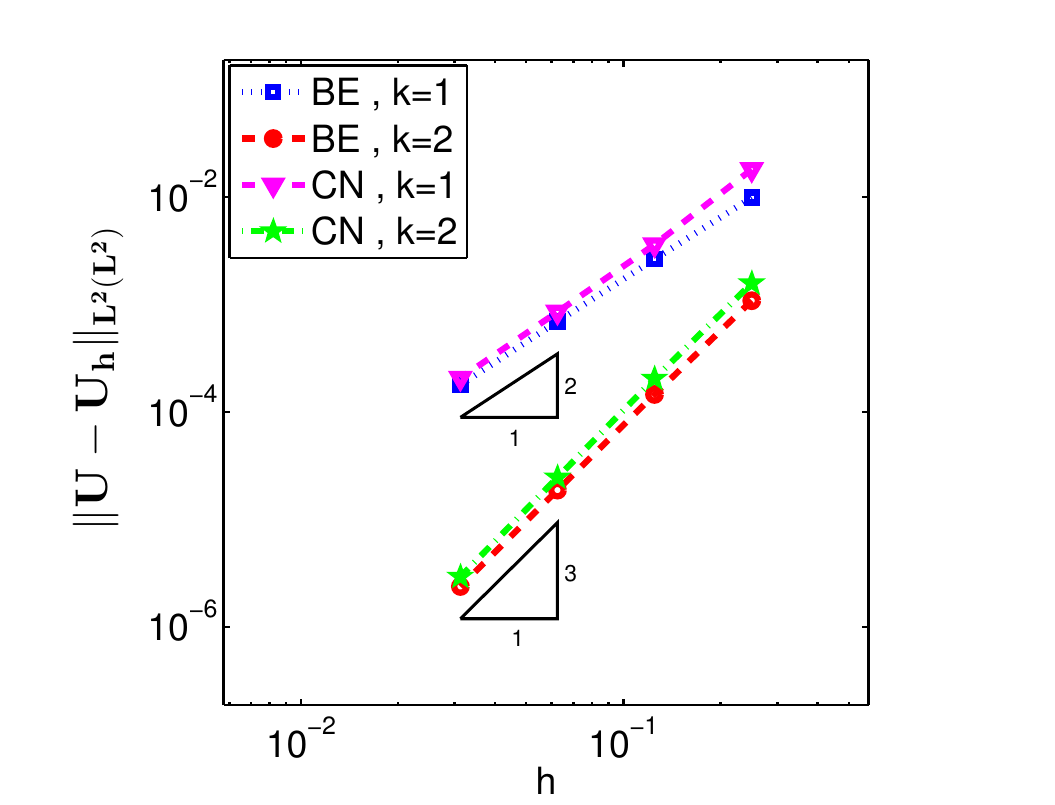}}
	\subfloat{\includegraphics[width=0.53\linewidth]{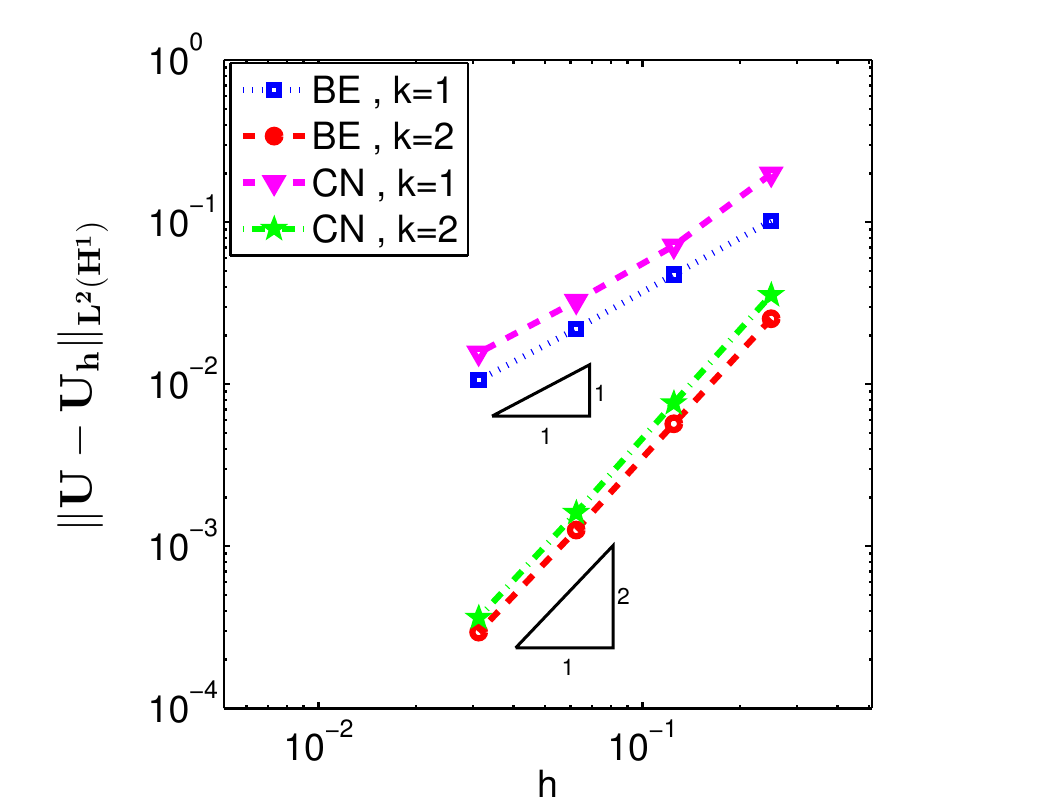}}
	\caption{Numerical orders of convergence in $L^2(L^2)$ (left) and in $L^2(H^1)$ (right) norm}
	\label{error}
\end{figure}

Theoretically, the expected convergence rates for SIPG method in $L^2(L^2)$ norm with piecewise discontinuous polynomials of degree $k$ are ${\mathcal O} (\Delta \tau + h^{k+1})$ with BE  and ${\mathcal O} (\Delta \tau^2 + h^{k+1})$ with CN, whereas they are ${\mathcal O} (\Delta \tau + h^{k})$ with BE  and ${\mathcal O} (\Delta \tau^2 + h^{k})$ with CN
in $L^2(H^1)$  norm \cite{riviere08dgm}. Since the convergence orders of BE and CN time integrators are well-known, we simultaneously compute the convergence orders with respect to space and time as in \cite{chen14haf}. We choose time step size $\Delta \tau = h^{k+1}$  with BE time integrator, and  $\Delta \tau = h^{(k+1)/2}$ with CN time integrator so that the global orders are given according to the space discretization as $ {\mathcal O} (h^{k+1})$ in $L^2(L^2)$ norm and $ {\mathcal O} (h^{k})$ in $L^2(H^1)$ norm. More precisely, with the BE method we set $\Delta \tau=h^2$ for linear polynomials and $\Delta \tau=h^3$ for  quadratic polynomials, while with the CN method we set  $\Delta \tau=h$ for linear polynomials and $\Delta \tau=h^{3/2}$ for quadratic polynomials. The expected orders of convergence are attained by the SIPG method in Fig.~\ref{error}. These a priori orders of convergence are optimal for the SIPG, and suboptimal for the two other variants of the dGFEM, i.e., the NIPG and IIPG methods \cite{Sun05}.

\subsection{European call option with Dirichlet boundary conditions}
\label{testex2}

To illustrate that SIPG  provides accurate solutions for the Heston model, we consider
the valuation of European call options due to the fact that its semi-analytical benchmark  price can easily be derived \cite{Heston93}.
Let $U(\tau,v,x)$ be the price of a European call option at time $\tau$,  with its payoff function
\begin{equation*}
	U^0(v,x)= (Ke^x-K)^{+},
\end{equation*}
where $K$ being the strike price of the option. We consider the RCD \eqref{convdiff_x} under the following Dirichlet type boundary conditions \cite{chen14haf}
	\begin{align*}
		U(\tau,v_{\text{min}},x)&=(Ke^{x-r_f\tau}-Ke^{-r_d\tau})^{+},\\
		U(\tau,v_{\text{max}},x)&=Ke^{x-r_f\tau},\\
		U(\tau,v,x_{\text{min}})&=0, \\
		U(\tau,v,x_{\text{max}})&=(Ke^{x_{\text{max}}-r_f\tau}-Ke^{-r_d\tau})^{+}.
	\end{align*}
We consider again the bounded spatial domain $\Omega=[0,4]\times [-2,2]$ with the mesh size $\Delta v = \Delta x = 0.0625$, and the time step size is taken as
$\Delta \tau=0.01$. The values of the system parameters are the same as in Example~\ref{testex1}.

In Table~\ref{relativeerror_call}, we compare the prices at the final time $\tau =1$ obtained by the SIPG using linear and quadratic elements with the semi-analytical solutions (exact prices) in \cite{Heston93}, at $(v_0,x_0)=(0.25,\log(S_0/K))$ with $S_0=100$.
The relative error is defined as
$
|\text{price-exact price}|/|\text{exact price}|
$.
The SIPG with linear and quadratic elements turns out to provide highly accurate prices for the call option with different strike prices.

\begin{table}[htb!]
	\centering
	\caption{Comparison with the closed form solutions for different strike prices in Example~\ref{testex2}: linear (quadratic) dG elements}
	\label{relativeerror_call}
	\begin{tabular}{lrcc}
		\hline\noalign{\smallskip}
		$K$& exact price & relative error \\ [2ex]
		\noalign{\smallskip}\hline\noalign{\smallskip}
		90  &  23.464  &  4.48e-04 (4.73e-05) \\
		95  &  20.739  &  2.75e-04 (5.12e-05) \\
		100 &  18.231  &  1.60e-03 (1.59e-05) \\
		105 &  15.938  &  1.79e-04 (5.33e-05) \\
		110 &  13.857  &  1.79e-03 (5.25e-05) \\
		115 &  11.979  &  5.16e-04 (1.26e-04) \\
		130 &  7.483   &  1.56e-03 (2.05e-04) \\
		150 &  3.701   &  5.42e-04 (1.99e-04) \\
		\noalign{\smallskip}\hline
	\end{tabular}
\end{table}

\subsection{Convection  dominated  European call option under Heston model}
\label{testex3}

We consider the convection dominated case for European call options \cite{Kluge02}. In order to illustrate the effect of large convective terms in the context of option pricing, we present numerical results for the high foreign interest rates. In the literature, for convection dominated option pricing models, usually, special predefined nonuniform grids are used for FDMs \cite{During14,Hout10adi,Kluge02}.

There are two critical points for the European call options under Heston model. The first one is around $S = K$ where the option is at the money. The other one is near the boundary $v = 0$, at which the oscillations occur due to the large convection coefficients relative to the diffusion coefficients,  making the PDE convection dominated in this region. The solution to the evolution problems modeled by the convection dominated RCD equations has a number of challenges. On one hand, one has to resolve the solution around the interior/boundary layers due to convection domination. On the other hand, the nature of nonstationary model leads to the resolution of spatial layers to be more critical since the location of the layers may vary as time progresses. In the case of convection dominated Heston model, the location of the layer is not changed as time progresses. Therefore the adaptive grid is constructed at the beginning of the time integration for the stationary RCD equation and it is used in all the succeeding time steps.

The adaptive dGFEM consists of finding a nonuniform mesh $\xi_h:=\xi_h^{(s)}$ ($s>0$) starting from a coarse uniform mesh $\xi_h^{(0)}$ by successive loops of the following  sequence
$$
\mathrm{SOLVE} \longrightarrow \mathrm{ESTIMATE} \longrightarrow \mathrm{MARK} \longrightarrow \mathrm{REFINE}
$$

Thus, on the $s$-th iteration, we solve the system \eqref{fullydisc} for the vector $\bm{u}^1$ and we obtain the solution $U_h^1$ on the mesh $\xi_{h}^{(s-1)}$. Using computed solution $U_h^1$, then, the local error indicators are calculated on each triangle $K\in\xi_{h}^{(s-1)}$, and according to local error indicators, the elements having large error are refined to obtain the new nonuniform mesh $\xi_{h}^{(s)}$. Here, the key step is the estimation of the local error indicators by the use of the only computed solutions and given problem data (a posteriori). As  the a posteriori error indicator, we modify the robust (independent of the P\'{e}clet number)  residual based error indicator in \cite{Schotzau09}, derived for a linear stationary RCD equation. In \cite{Schotzau09}, the diffusion term is a constant scalar, but here the diffusion term is a nonconstant matrix. So, our modification is mainly related to the imposing of the diffusion matrix $A$ into the formulations.  Let assume that there are nonnegative real numbers $r_*$ and $c_*$ satisfying the following conditions:
	\begin{align*}
		r_d - \frac{1}{2}\nabla \cdot \mathbf{b} &\geq r_* ,\\
		\| r_d - \nabla \cdot \mathbf{b} \|_{L^{\infty}(\Omega)} &\leq c_*r_*.
	\end{align*}
Indeed, the first identity above is necessary for the well-posedness of the problem in PDE form, and the second one is required for the reliability of the proposed a posteriori error estimator.
In order to measure the local errors
for each element $K \in \xi_{h}^{(s-1)}$, we define the local error indicators $\eta_K^2$:
\begin{equation*}
	\eta_K^2= \eta_{R_K}^2  + \eta_{E_K^0}^2 + \eta_{E_K^D}^2+ \eta_{E_K^N}^2,
\end{equation*}
where $\eta_{R_K}$ denotes the element residuals given by
\begin{equation*}
	\eta_{R_K}^2 = \omega_K^2\| (U_h^1-U_h^{0})/\Delta\tau - \nabla\cdot(A(v)\nabla U_h^1)+\mathbf{b}\cdot\nabla U_h^1+r_dU_h^1\|_{L^2(K)}^2,
\end{equation*}
while, $\eta_{E_K^0}$, $\eta_{E_K^D}$ and $\eta_{E_K^N}$ stand for the edge residuals coming from the jump of the numerical solution on the interior, Dirichlet boundary and Neumann boundary edges, respectively, given by
\begin{align*}
	\eta_{E_K^0}^2 &= \sum \limits_{e \in \partial K\cap\Gamma_h^{0}}\left(\frac{1}{2}\epsilon_K^{-\frac{1}{2}}\omega_e\left\| \jump{A\nabla U_h^1}_e\right\|_{L^2(e)}^2+\frac{1}{2}\left(\frac{\epsilon_K\gamma}{h_e}+r_* h_e+\frac{h_e}{\epsilon_K}\right)\left\| \jump{U_h^1}_e\right\|_{L^2(e)}^2\right), \\
	\eta_{E_K^D}^2 &= \sum \limits_{e \in \partial K\cap\Gamma_h^{D}}\left(\frac{\epsilon_K\gamma}{h_e}+r_* h_e+\frac{h_e}{\epsilon_K}\right)\left\| U^D(\tau^1,\mathbf{z})-U_h^1\right\|_{L^2(e)}^2,  \\
	\eta_{E_K^N}^2 &= \sum \limits_{e \in \partial K\cap\Gamma_h^{N}}\epsilon_K^{-\frac{1}{2}}\omega_e\left\| U^N(\tau^1,\mathbf{z})-A\nabla U_h^1\cdot \mathbf{n}_K\right\|_{L^2(e)}^2.
\end{align*}
In the above formulation, $\omega_K$ and $\omega_e$ are positive weights given by
$$
\omega_K = \min\{h_K\epsilon_K^{-\frac{1}{2}},r^{-\frac{1}{2}}_*\} \; , \quad  \omega_e = \min\{h_e\epsilon_K^{-\frac{1}{2}}, r^{-\frac{1}{2}}_* \},
$$
where $h_K$ and $h_e$ denote the sizes of the element $K$ and the edge $e$, respectively. The global a posteriori error indicator is then given by
$$
\eta=\left( \sum \limits_{K\in{\xi}_{h}^{(s-1)}}\eta_K^2\right)^{1/2}.
$$
The reliability and efficiency proofs followed by a similar procedure in \cite{Schotzau09} (see \cite{Schotzau09} for details).
The iteration continues until the error indicator $\eta$ satisfies a prescribed tolerance, producing the fix nonuniform mesh ${\xi}_{h}:={\xi}_{h}^{(s)}$ to be used in the successive time steps, for some $s\in\mathbb{Z}^+$.

\begin{table}[htb!]
	\centering
	\caption{ Parameter set for Example~\ref{testex3}}
	\label{convection_dominated_parameters}
	\resizebox{\columnwidth}{!}{%
	\begin{tabular}{cccccccc}
		\hline\noalign{\smallskip}
		$\kappa$& $\theta$ & $\sigma$& $\rho$ & $r_d$  & $r_f$& $T$&$K$  \\
		\noalign{\smallskip}\hline\noalign{\smallskip}
		1.98937 & 0.011876 & 0.33147 & 0.0258519 & $\log(1.0005)$ & {$\log(100)$}  &0.25&123.4  \\
		\noalign{\smallskip}\hline
	\end{tabular}
	}
\end{table}

For the numerical experiments, we impose the following initial-boundary conditions proposed in~\cite{Winkler01}
\begin{align*}
	U(\tau,v_{\text{min}},x)&=Ke^{x-r_f\tau}\Phi(d_{+})-Ke^{r_d\tau}\Phi(d_{-}), \\
	U(\tau,v_{\text{max}},x)&=Ke^{x-r_f\tau}, \\
	U(\tau,v,x_{\text{min}})&=\lambda U(\tau,v_{\text{max}},x_{\text{min}})+(1-\lambda)U(\tau,v_{\text{min}},x_{\text{min}}), \\
\lambda &= \frac{v-v_{\text{min}}}{v_{\text{max}}-v_{\text{min}}}, \\
	A\nabla U(\tau,v,x_{\text{max}})\cdot\mathbf{n} &=\frac{1}{2}v K e^{x-r_f\tau}, \\
	U(0,v,x)&=(Ke^x-K)^{+} ,
\end{align*}
where
\begin{equation*}
	d_{+}=\frac{x+\left(r_d-r_f+\frac{1}{2}v_{\text{min}}\right)\tau}{\sqrt{v_{\text{min}}\tau}},\quad d_{-}=\frac{x+\left(r_d-r_f-\frac{1}{2}v_{\text{max}}\right)\tau}{\sqrt{v_{\text{max}}\tau}},
\end{equation*}
and $\Phi(x)$ is the cumulative distribution function given as
\begin{equation*}
	\Phi(x)=\frac{1}{2\pi}\int_{-\infty}^{x}e^{-y^2/2}dy.
\end{equation*}

The parameters are taken from \cite{Kluge02} (see Table~\ref{convection_dominated_parameters}).  We consider a bounded domain $\Omega =(0.0025,0.559951)\times(2.990790,6.640072)$ for the computational purpose. The constant time step size is taken as $\Delta\tau=0.0125$. Precisely, we choose  $r_f=\log(100)$ to examine the effect of large convective term in a superior way.

\begin{figure}[htb!]
	\centering
	\subfloat{\includegraphics[width=0.5\linewidth]{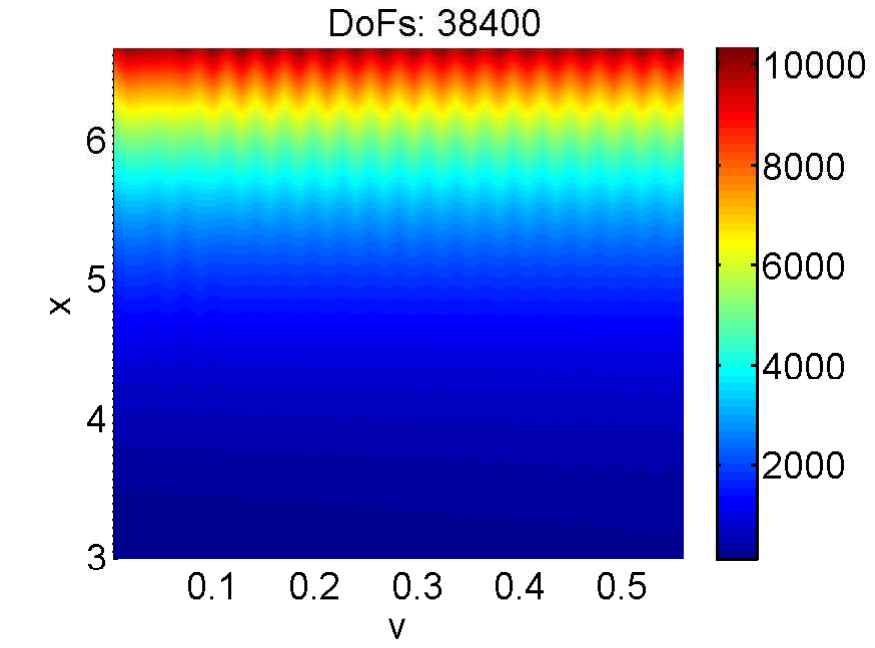}}
	\subfloat{\includegraphics[width=0.5\linewidth]{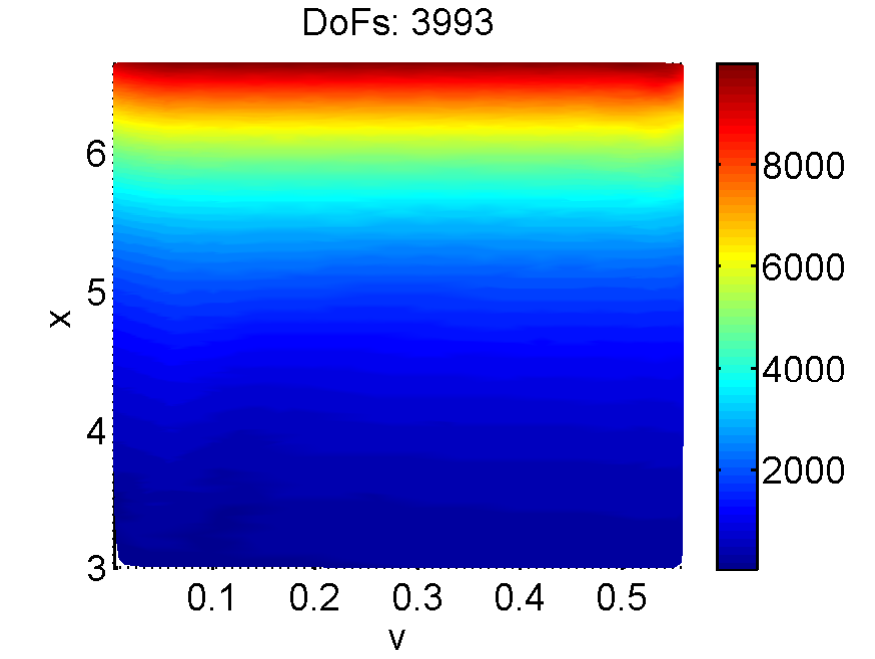}}
	\caption{Solution profiles in Example~\ref{testex3} at $\tau=0.25$ on the  uniform (left) and adaptive (right) mesh \label{adap1}}
\end{figure}

\begin{figure}[htb!]
	\centering
	\includegraphics[width=0.5\linewidth]{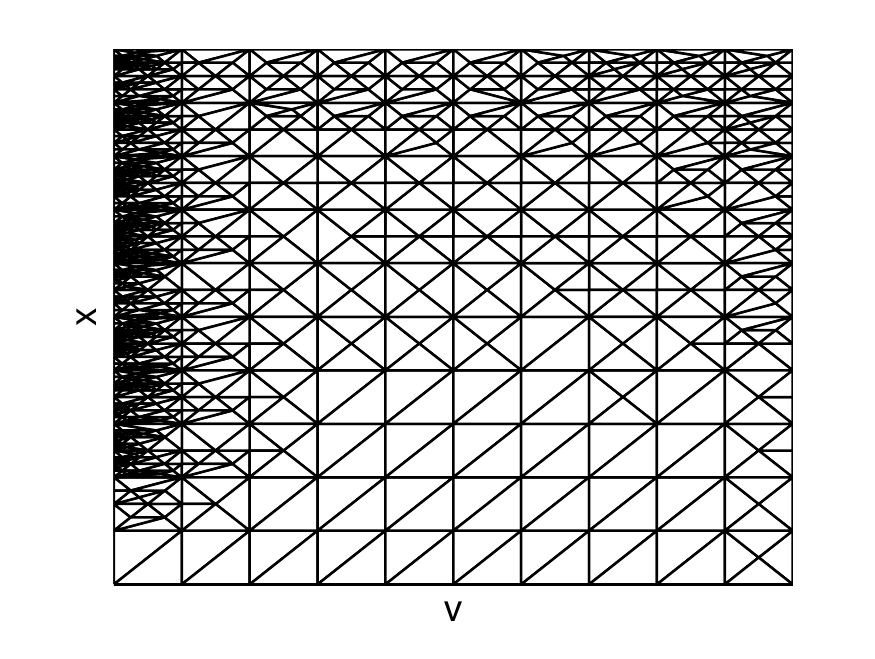}
	\caption{Adaptive mesh in Example~\ref{testex3} \label{adap2}}
\end{figure}

The adaptive solution in Fig.~\ref{adap1}, right, produces accurate solutions without oscillations using fewer degrees of freedom (DoFs) (one-tenth) than the oscillating solutions on the uniform meshes, Fig.~\ref{adap1}, left. The boundary layers are also accurately detected by the adaptive algorithm as shown in Fig.~\ref{adap2}.

\subsection{Digital call option}
\label{testex4}

We examine the performance of  SIPG with Rannacher smoothing for the digital call options with a discontinuous payoff \cite{england06,Lazar03}
\begin{equation*}
	U^0(v,x)= \mathbbm{1}_{\{Ke^x>K\}},
\end{equation*}
where $K$ is the strike price of the option, which is treated as a barrier level. Precisely, if the underlying price is at or below the level $K$ at maturity, the option will be worthless; otherwise it will pay 1 unit of money at time $T$. Let $U(\tau,v,x)$ be the price of a digital call option adhering  RCD equation \eqref{convdiff_x}.  Inspiring by~\cite{england06}, we impose the following  Dirichlet type boundary conditions in the $x$-direction
\begin{equation*}
	U(\tau,v,x_{\text{min}})=0, \quad U(\tau,v,x_{\text{max}})=e^{x_{\text{max}}-r_f\tau}.
\end{equation*}
The idea behind this boundary condition can be given as follows:  when the price of the underlying security is very  high as compared to the strike price, then the option is worth $e^{x-r_f\tau}$ as reaching to a value of 1 at maturity. Moreover,  we prescribe the following boundary conditions in the $v$-direction~\cite{england06}
$$
\frac{\partial U}{\partial v}(\tau,v_{\text{min}},x) = 0, \quad \frac{\partial U}{\partial v}(\tau,v_{\text{max}},x) = 0.
$$
Indeed, when $v\rightarrow 0$ and $v\rightarrow\infty,$ it is observed that the price of the underlying security becomes steady~\cite{england06}. As a result, the value of the option does not show sensitivity for the extreme values of volatility.

The numerical computations are performed on the  bounded domain $\Omega =(0.0025,0.559951)\times(-5,5)$ with the mesh size $\Delta v = 0.016$, $\Delta x = 0.078$, and the time step size $\Delta\tau= 0.025$. Parameter set is
taken from \cite{Winkler01} (see Table~\ref{Neumann_parameters}).

\begin{table}[htb!]
	\centering
	\caption{Parameter set for Example~\ref{testex4}}
	\label{Neumann_parameters}
	\begin{tabular}{cccccccc}
		\hline\noalign{\smallskip}
		$\kappa$& $\theta$ & $\sigma$& $\rho$ & $r_d$ & $r_f$  & $T$&$K$  \\
		\noalign{\smallskip}\hline\noalign{\smallskip}
		2.5 & 0.06 & 0.5 & -0.1 & $\log(1.052)$ & $\log(1.048)$ & 0.25&1  \\
		\noalign{\smallskip}\hline
	\end{tabular}
\end{table}

\begin{figure}[htb!]
	\centering
	\subfloat{\includegraphics[width=0.5\linewidth]{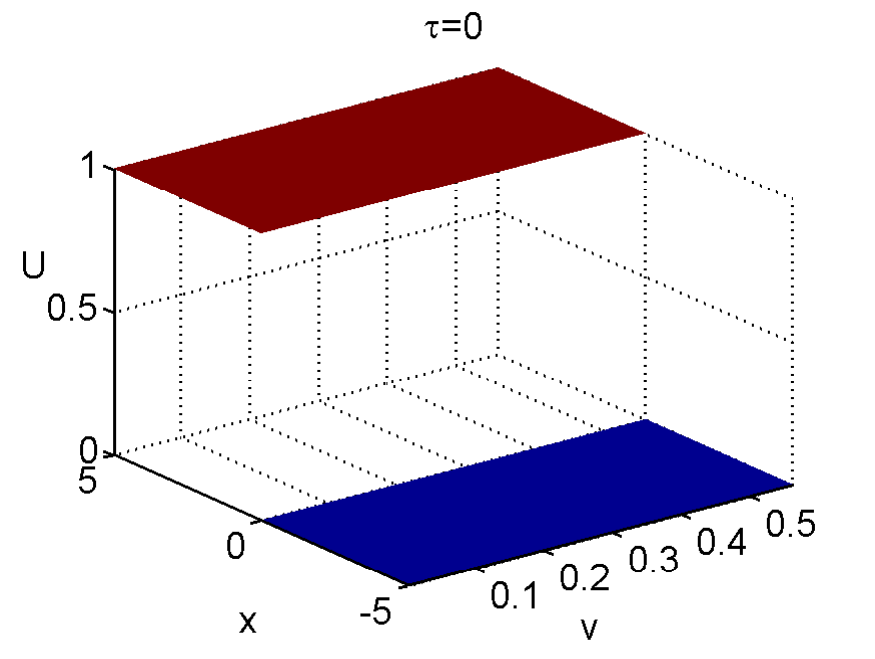}}
	\subfloat{\includegraphics[width=0.5\linewidth]{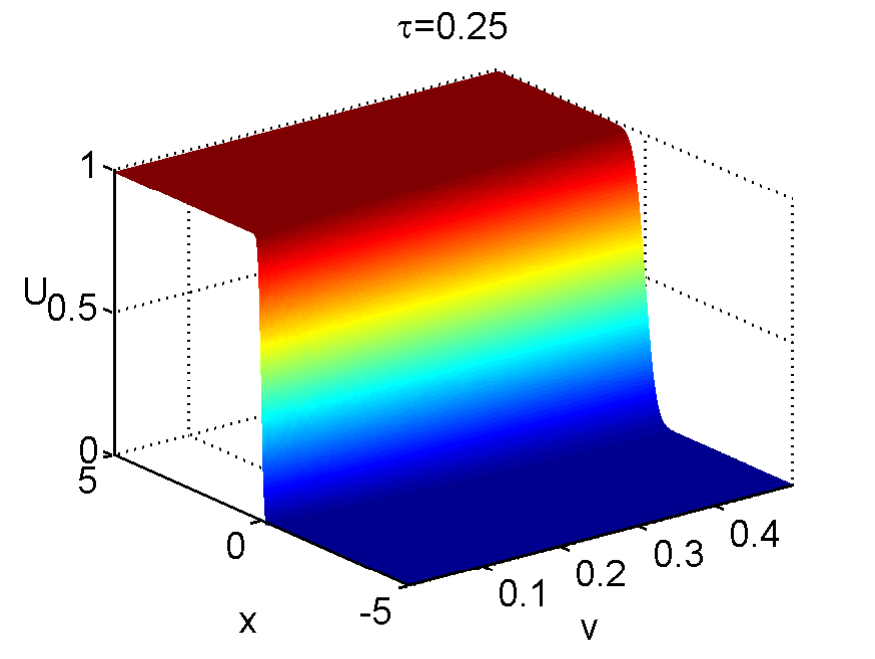}}
	\caption{Price profiles of the European digital call option in Example~\ref{testex4} at  $\tau=0$ (left) and at $\tau=0.25$ (right)}
	\label{numexWinkler_digital}
\end{figure}

Fig.~\ref{numexWinkler_digital} illustrates the payoff (left)  and price  surface (right) of the  digital call option. The payoff function is discontinuous  at $x=0$, whereas the price is smoother as a result of the diffusion effect. The effect of the diffusion term is more visible for large volatilities.

In Table~\ref{relerror_BECNRannacher}, we present the relative errors
$|\text{price}-\text{price}_{\text{ref}}|/|\text{price}_{\text{ref}}|$ at $(v_0,x_0)=(0.05225,0)$,  and at the final time $\tau=0.25$ obtained by the CN method  with and without Rannacher smoothing for linear dG elements. The reference solution $\text{price}_{\text{ref}}=0.483827$ is obtained by the semi-analytical formula in \cite{Lazar03}.
In the table,  $N_v$ and $N_x$ denote the number of partitions in $v$ and $x$-directions, respectively. It is apparent that the solutions obtained by Rannacher smoothing produce by far more accurate solutions than the solutions obtained by CN method, especially on fine grids.

\begin{table}[htb!]
	\centering
	\caption{Relative errors for digital call option in Example~\ref{testex4}}
	\label{relerror_BECNRannacher}
	\begin{tabular}{l l  c  c c  c}
		\hline\noalign{\smallskip}	
		\multicolumn{2}{l}{} &  \multicolumn{2}{c}{CN } & \multicolumn{2}{c}{CN with Rannacher smoothing}\\
		\noalign{\smallskip}\hline\noalign{\smallskip}
		$N_v$&$N_x$  &  price & relative error  & price & relative error  \\
		\noalign{\smallskip}\hline\noalign{\smallskip}
		8 & 16 &  0.524935 &8.50e-02    & 0.524910 &  8.49e-02 \\
		16 & 64 &  0.494234 &  2.15e-02  & 0.496226 &  2.56e-02  \\
		32 & 128 &  0.368798 & 2.38e-01   & 0.484065
		& 4.93e-04\\
		64 & 256 & 0.554879 & 1.47e-01   & 0.483568 & 5.34e-04\\
		\noalign{\smallskip}\hline
	\end{tabular}
\end{table}

\subsection{American put option}
\label{testex5}

We present numerical results for American put options solved by the preconditioned PSOR method (Algorithm~\ref{psor}) with Rannacher smoothing. The payoff function of American option is  given by
\begin{equation*}
	U^0(v,x)=(K-Ke^x)^+,
\end{equation*}
where  $K$ is the strike price of the option.  Let $U(\tau,v,x)$ denote the option price satisfying the inequality system \eqref{pdeA_American} on the bounded domain $\Omega =(0.0025,0.5)\times(-5,5)$. We consider the following  boundary conditions \cite{kunoth2012multiscale,Zhu09}
$$
U(\tau,v,x_{\text{min}})=K, \quad U(\tau,v,x_{\text{max}})=0,
$$
$$
U(\tau,v_{\text{min}},x) = (K-Ke^x)^+\; , \quad
\frac{\partial U}{\partial v}(\tau,v_{\text{max}},x) = 0.
$$
We use the uniform time step size $\Delta \tau = 0.01$, and a graded grid in the direction of the transformed
underlying price $x$, which is four times as large as the grid for the variance $v$, like in \cite{kunoth2012multiscale}.

For evaluation of the accuracy and the performance of the preconditioned PSOR method, we take the parameter values given in Table~\ref{par:American}.
The results are given in Table~\ref{psornum}, where the prices are evaluated at the final time $\tau =0.25$ and at $(v_0,x_0)=(0.25,0)$. Table~\ref{psornum} shows that an average number of the PSOR iterations are almost the same and independent of the space discretization for each preconditioner, which is characteristic for many iterative methods like the multigrid methods  \cite{clarke1999multigrid,kunoth2012multiscale,oosterlee2003multigrid}. The CPU times (in seconds) are larger for the left-right preconditioner due to the matrix square function ${\bf sqrtm}$ of MATLAB. In  \cite{kunoth2012multiscale} option prices at the final time $\tau =0.25$ and at $(v_0,x_0)=(0.25,0)$ are listed for the same problem which varies between $0.75-0.80$  for different methods.  Among the three preconditioners the left-right preconditioner produces more stable values of the prices with finer grids and the produced values are more close to those in the literature \cite{kunoth2012multiscale}. Therefore, in the following computations, we use the left-right preconditioner on a $20 \times 80$ grid, which also close to the approximate prices of $0.794969$ and $0.795687$ obtained by Gauss-Seidel and monotone multigrid methods, respectively.

\begin{table}[htb]
	\centering
	\caption{ Parameter set for American put option in Example~\ref{testex5}}
	\label{par:American}
	\begin{tabular}{cccccccc}
		\hline\noalign{\smallskip}
		$\kappa$ & $\theta$ & $\sigma$ & $\rho$ & $r_d$ & $r_f$ & $T$ & $K$  \\
		\noalign{\smallskip}\hline\noalign{\smallskip}
		5   &   0.16   &   0.9   & 0.1 &  0.1  & 0  & 0.25  & 10  \\
		\noalign{\smallskip}\hline
	\end{tabular}
\end{table}

\begin{table}[htb!]
	\centering
	\caption{PSOR algorithm results for American put option in  Example~\ref{testex5}}
	\label{psornum}
	\resizebox{\columnwidth}{!}{%
	\begin{tabular}{cc|ccc|ccc|ccc}\hline
		&       &      \multicolumn{3}{c|}{left preconditioner} & \multicolumn{3}{c|}{right preconditioner} & \multicolumn{3}{c}{left-right preconditioner} \\ \hline
		$N_v$ & $N_x$ & price      & CPU time     & \# iter     & price       & CPU time      & \#iter     & price         & CPU time       & \#iter       \\\hline
		8     & 32     & 0.9090     & 2.2          & 7.7         & 0.7950      & 1.9           & 6          & 0.7791        & 3.0            & 6.6          \\
		12    & 48    & 0.8475     & 9.5          & 8           & 0.8342      & 7.7           & 6          & 0.8024        & 19.4           & 7            \\
		16    & 64    & 0.8257     & 30.4         & 8           & 0.8350      & 24.3          & 6          & 0.8053        & 95.7           & 7            \\
		20    & 80    & 0.8138     & 73.5         & 8           & 0.8346      & 61.3          & 6          & 0.8042        & 307.7          & 7 \\ \hline
	\end{tabular}%
}
\end{table}

\begin{figure}[htb!]
	\centering
	\subfloat{\includegraphics[width=0.5\linewidth]{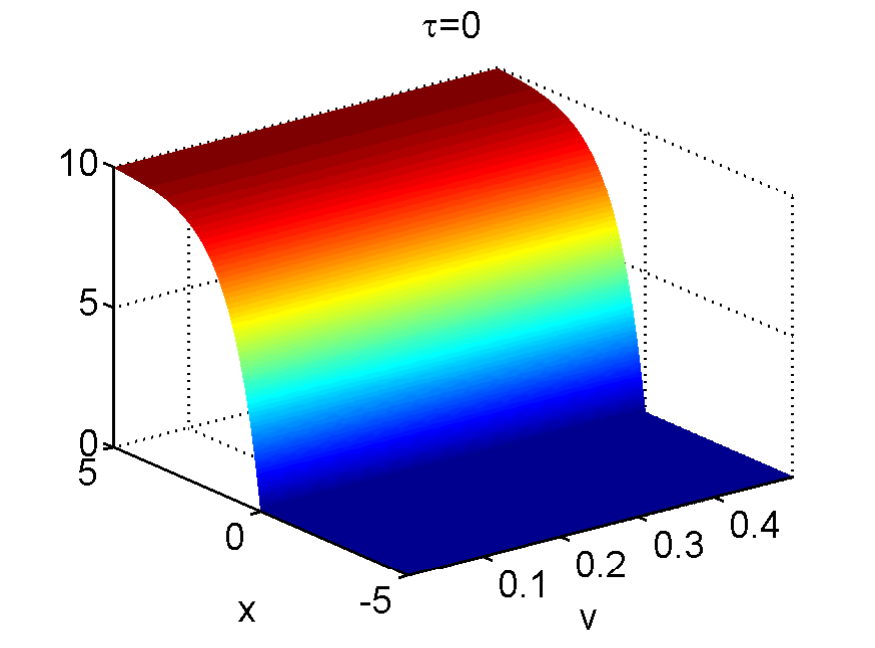}}
	\subfloat{\includegraphics[width=0.5\linewidth]{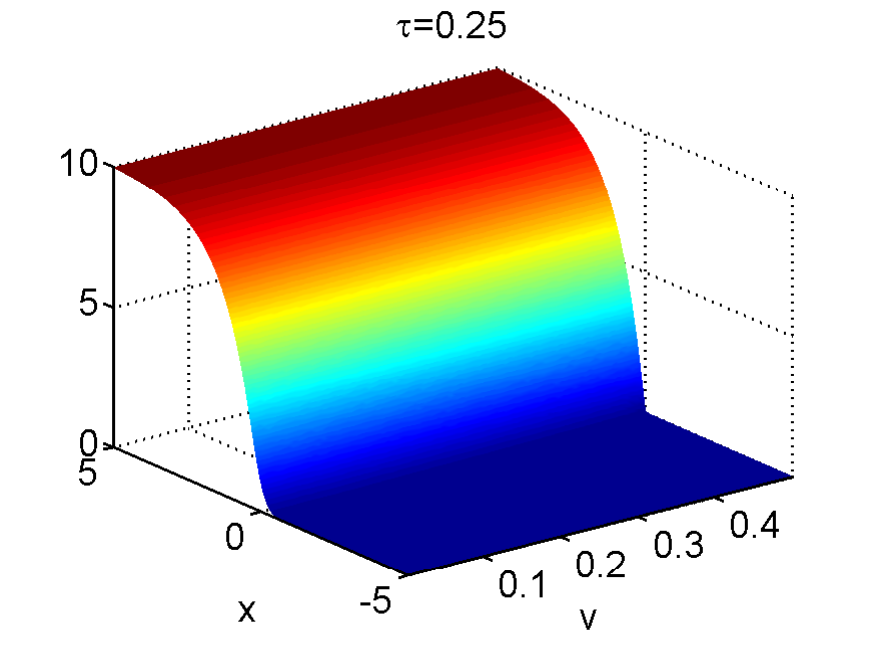}}
	\caption{Price profile of American put option in Example~\ref{testex5} at $\tau=0$  (left) and $\tau=0.25$ (right) \label{fig:American1}}
\end{figure}

\begin{figure}[htb!]
	\centering
	\subfloat{\includegraphics[width=0.5\linewidth]{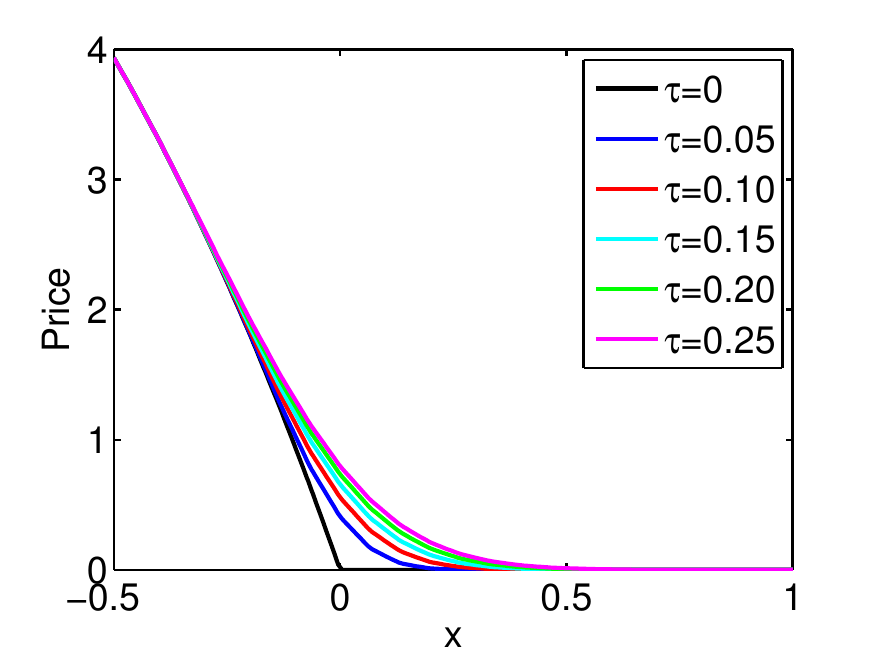}}
	\caption{Evolution of the option price in Example~\ref{testex5} at different times $\tau$ with constant variance $v = 0.25$ \label{fig:American2}}
\end{figure}

The full price surfaces at times $\tau=0$\ and $\tau=0.25$, and
option prices at different times $\tau$ for constant variance $v = 0.25$ are shown in Fig.~\ref{fig:American1} and Fig.~\ref{fig:American2}, respectively. The characteristic bend of the termination condition in the American option  is smoothed out, as can be seen in Fig.~\ref{fig:American1}.  This is a consequence of the parabolic behavior of the solution operator, which is observed by all numerical methods for the American options (see for example \cite{kunoth2012multiscale}). In Fig.~\ref{fig:American2}, the lowest plot  shows the payoff function $U^0(v,x)$, and  it is clear that  the  bend is smoothed as time goes  from $\tau=0$  to $\tau=0.25$. We can also notice that at different time instances the option price is projected to the payoff function $U^0(v,x)$.

Moreover,  in the following test example, we compute option prices for  different values of the volatility-of-variance parameter $\rho \in [ -1,1]$ which defines the correlation between the two underlying Brownian motions in  Heston model.  We consider that the Feller condition $ 2\kappa \theta \geq \sigma^2$ is fulfilled. In  Table~\ref{par:American2}, the parameters are given for the case when the Feller condition is satisfied \cite{Hout15}.

\begin{table}[htbp]
	\centering
	\caption{ Parameter set when Feller condition is satisfied}
	\label{par:American2}
	\begin{tabular}{ccccccc}
		\hline\noalign{\smallskip}
		$\kappa$ & $\theta$ & $\sigma$ & $r_d$ & $r_f$ & $T$ & $K$   \\
		\noalign{\smallskip}\hline\noalign{\smallskip}
		0.6067       &   0.0707       &   0.2928     &    0.03     &  0     & 3     & 100  \\
		\noalign{\smallskip}\hline
	\end{tabular}
\end{table}

In Table~\ref{tab:corr} the prices evaluated at $(v_0,x_0)=(0.05, \log(S_0/K)$ and at the final time $\tau=3$ are listed for different correlations  based on the parameter set given in Table~\ref{par:American2}, for which the Feller condition is satisfied. In \cite{Hout15}, a detailed comparison of American option prices is given for different correlations by using various methods. In Table~\ref{tab:corr}, the prices obtained in \cite{Hout15} are listed in parenthesis  for $\rho=-0.7571$.  Because the boundary conditions in \cite{Hout15} are different,  a direct comparison with our results is not possible.  But from Table~\ref{tab:corr} it can be seen that the option prices are decreasing with increasing $S_0$, as in \cite{Hout15}.

\begin{table}[htbp]
	\centering
	\caption{American put option prices in Example~\ref{testex5} for different correlations}
	\label{tab:corr}
	\begin{tabular}{lrrrrr}
		\hline\noalign{\smallskip}
		& $\rho=-0.7571$ & $\rho=-0.5$ & $\rho=0$ & $\rho=0.5$ & $\rho=0.7571$ \\
		\noalign{\smallskip}\hline\noalign{\smallskip}
		$S_0=90$  &  14.9206 (16.0470)   & 15.4264   & 16.3747  &   17.2760  &  17.7561 \\ [1ex]
		$S_0=100$ &  10.4266 (12.4326)   & 10.7104   & 11.1430  &   11.5011  &  11.7019 \\[1ex]
		$S_0=110$ &  8.5346 (9.8746)    & 8.4332    & 8.0937   &    7.5108  &  7.0502 \\
		\noalign{\smallskip}\hline
	\end{tabular}
\end{table}

\subsection{Comparison of dGFEM with radial basis functions}
\label{testex6}

We compare the dGFEM with the RBF-PUM  \cite{Mollapourasl19} for  European and American put options. In Table~\ref{par:rbf1}, we give the parameter set as in \cite{Mollapourasl19}, which is chosen for the case when the Feller condition is violated.

\begin{table}[htbp]
	\centering
	\caption{Parameter set for European and American put options in Example~\ref{testex6}}
	\label{par:rbf1}
	\begin{tabular}{cccccccc}
		\hline\noalign{\smallskip}
	$\kappa$ & $\theta$ & $\sigma$ &   $\rho$ & $r_d$ & $r_f$ & $T$ & $K$  \\
		\noalign{\smallskip}\hline\noalign{\smallskip}
1.15   &   0.0348      &   0.39   & -0.64   &    0.04     &  0     & 0.25     & 100  \\
		\noalign{\smallskip}\hline
	\end{tabular}
\end{table}

We impose the following boundary  and initial conditions
\begin{itemize}
	\item  For European put option
	\begin{align*}
		& U(\tau,v,x_{\text{min}})= Ke^{-r_d\tau}, \qquad U(\tau,v,x_{\text{max}})= 0, \\
		& \frac{\partial U}{\partial v}(\tau,v_{\text{max}},x) = 0, \qquad U(\tau,v_{\text {min}},x) = (Ke^{-r_d\tau}-Ke^x)^+,\\
		& U(0,v,x)= (K-Ke^x)^+.
	\end{align*}
	\item For American put option
	 \begin{align*}
	& U(\tau,v,x_{\text{min}})=K,\qquad U(\tau,v,x_{\text{max}})=0,\\
	& \frac{\partial U}{\partial v}(\tau,v_{\text{max}},x) = 0,\qquad
		U(\tau,v_{\text{min}},x) = (K-Ke^x)^+,\\
		& U(0,v,x)= (K-Ke^x)^+,
	\end{align*}
\end{itemize}
on the bounded domain $\Omega =(0,0.5)\times (-\log(2),\log(2))$,
where the region of interest for the underlying price then becomes $(K/2, 2K)$.

In order to check the accuracy of SIPG (NIPG) results, we use the following averaged error in \cite{Mollapourasl19}
\begin{equation*}
E_{\text{avg}}(x_1,\ldots , x_N) = \sqrt{\frac{1}{N} \sum_{i=1}^N (U(T,v_0,x_i) -U_{\text{ref}}(T,v_0,x_i))^2},
\end{equation*}
for a set of points $x_i=\log(S_i/K)$, $i=1,\ldots ,N$, where $U_{\text{ref}}$ stands for a reference price function which is taken as the exact solution given in \cite{Mollapourasl19} for the European option, and the reference solutions reported in \cite{Hout15} and \cite{oosterlee11} for the American option. Here we use $N=3$ points with $S_1=90$, $S_2=100$ and $S_3=110$.
In addition, we use less spatial grid points in the $v$-direction than in the $x$-direction to increase the efficiency (see for a similar approach in  \cite{Hout15}).
In \cite{Mollapourasl19} for RBF-PUM,  patches are needed to cover the domain. The sparsity pattern of the matrices in \cite{Mollapourasl19} depends on the number of patches and of nodes. For the same number of nodes, more patches lead to more sparsity.  For dG methods, the sparsity pattern depends only on the number of nodes, for linear elements it has a banded structure with a $3\times 3$ block structure. Therefore a direct comparison of the efficiency of the RBF-PUM with dG methods is not possible because of the different structures of both methods.

The results for $v_0=0.0348$ at the final time $\tau =0.25$ are presented in Table~\ref{sipg_european} and Table~\ref{sipg_american} for European and American options, respectively.
A large penalty parameter ($\gamma =50$) is taken for the NIPG.
In the case of NIPG for American options with quadratic elements, the errors are larger on  finer meshes due to convergence behavior of the iterative method, because of high condition numbers of the left-right preconditioned system matrix ${\tilde{\mathbf B}}$ of the LCP \eqref{lcp_American}, whereas for SIPG the condition numbers are small and independent of the mesh size.  
The error on the $12 \times 120$  grid in Table~\ref{sipg_european}  is close to the error value $2.4e-02$ in \cite{Mollapourasl19} on $38 \times 38 $ grid with $18 \times 18 $ patches. In the case of American option we obtained  by the SIPG method more accurate error $2.48e-02$ than for the error $1.22e-01$ by RBF-PUM in (Table 8, \cite{Mollapourasl19}) when the solutions in \cite{Hout15} are used as $U_{\text{ref}}$.  Similarly, it is smaller than the error $5.64e-02$ by RBF-PUM compared with the results of \cite{oosterlee11}.

\begin{table}[htb]
	\centering
	\caption{European put option prices in Example~\ref{testex6} with SIPG (NIPG) }
	\label{sipg_european}
	\resizebox{\columnwidth}{!}{%
	\begin{tabular}{lllcccc}
	\hline\noalign{\smallskip}
	&	$N_v$ & $N_x$   & $S_1=90$& $S_2=100$& $S_3=110$& $E_{\text{avg}}(x_1,x_2, x_3)$ \\
		\noalign{\smallskip}\hline\noalign{\smallskip}
\multirow{ 8}{*}{$k=1$} &	8 &	32 &	9.788 (9.785) &	2.243 (2.244) &	0.911 (0.910) &	5.68e-01 (5.66e-01) \\
  &	 8 &	64  &	 9.628 (9.627)  &	2.327 (2.327) &	0.900 (0.900)  &	4.89e-01 (4.89e-01) \\
  &	 8 &	72  &	 9.671 (9.670) &	2.339 (2.338) &	0.946 (0.945) &	 4.91e-01 (4.91e-01) \\
  &	8 &	 80 &	 9.623 (9.622) &	 2.348 (2.349) &	 0.907 (0.907)  &  4.76e-01 (4.76e-01) \\
 & 12 &   48   &  9.505 (9.503)   & 2.830 (2.823) &   0.969 (0.969) &   1.94e-01 (1.97e-01) \\
 & 12  &  96  &   9.491 (9.490)  &  2.873 (2.871)  &  0.959 (0.958)  &  1.67e-01 (1.68e-01) \\
 & 12 &  108   &  9.481 (9.480)  &  2.873 (2.872)   & 0.966 (0.966)  &  1.66e-01 (1.66e-01) \\
 & 12  & 120   &  9.478 (9.477)   & 2.871 (2.872)   & 0.971 (0.971)  &  1.66e-01 (1.66e-01)\\
 &	&  &  &   &    & \\
	\multirow{ 8}{*}{$k=2$} &	8  &  32  &   9.440 (9.439)  &  2.953 (2.948)  &  0.953 (0.952) &    1.13e-01 (1.16e-01) \\
 & 8  &  64  &   9.430 (9.429)  &  2.999 (2.998)  &  0.955 (0.955)  &  8.73e-02 (8.77e-02) \\
 & 8  &  72  &   9.422 (9.422)  &  3.008 (3.007)  &  0.956 (0.956)  &  8.15e-02 (8.16e-02) \\
 & 8  &  80  &   9.423 (9.423)  &  3.007 (3.008)   & 0.955 (0.955)  &  8.19e-02 (8.15e-02) \\
& 12  &  48  &   9.381 (9.381)  &  3.158 (3.156)   & 0.931 (0.931)  &  1.81e-02 (1.74e-02) \\
 &12  &  96  &   9.381 (9.381)  &  3.170 (3.169)  &  0.930 (0.930)  &  2.36e-02 (2.33e-02) \\
& 12  & 108   &  9.379 (9.379)  &  3.176 (3.175)  &  0.928 (0.929)   & 2.67e-02 (2.63e-02) \\
& 12  & 120  &   9.379 (9.379)  &  3.182 (3.181)  &  0.927 (0.927)  &  2.95e-02 (2.90e-02) \\
	 &	&  &  &   &    & \\
	\hline
\multicolumn{3}{c}{Exact prices} & 9.36868 & 3.13248 & 0.91752   \\
\noalign{\smallskip}\hline
	\end{tabular}
	}
\end{table}

\begin{table}[htb]
	\centering
	\caption{American put option prices in Example~\ref{testex6} with SIPG (NIPG) }
	\label{sipg_american}
	\resizebox{\columnwidth}{!}{%
	\begin{tabular}{lllcccc}
	\hline\noalign{\smallskip}
	&	$N_v$ & $N_x$   & $S_1=90$& $S_2=100$& $S_3=110$ & $E_{\text{avg}}(x_1,x_2, x_3)$ \\
		\noalign{\smallskip}\hline\noalign{\smallskip}
\multirow{ 8}{*}{$k=1$} &	8  &  32   &  10.466 (10.447)   & 2.457 (2.448)  &  0.933 (0.929) &    5.12e-01 (5.10e-01) \\
 & 8  &  64  &   10.282 (10.277) &   2.457 (2.458)  &  0.907 (0.906) &   4.64e-01 (4.63e-01) \\
 & 8  &  72  &   10.313 (10.309)  &  2.451 (2.454) &   0.952 (0.951) &   4.74e-01 (4.72e-01) \\
 & 8  &  80  &   10.271 (10.268)  &  2.476 (2.479) &   0.913 (0.912)  &  4.52e-01 (4.49e-01) \\
 &12  &  48   &  10.147 (10.123)  &  2.911 (2.901)  &  0.987 (0.983)  &  1.96e-01 (1.94e-01) \\
 &12 &   96  &   10.082 (10.079)  &  2.950 (2.951)  &  0.965 (0.965)  &  1.58e-01 (1.57e-01) \\
& 12  & 108  &   10.073 (10.071)  &  2.947 (2.948)  &  0.972 (0.971)  &  1.59e-01 (1.58e-01) \\
& 12 &  120  &   10.069 (10.067)  &  2.945 (2.947)  &  0.976 (0.976)  &  1.60e-01 (1.59e-01)\\
 &	&  &  &   &    & \\
	\multirow{ 8}{*}{$k=2$} &	8  &  32  &   10.017 (10.017)  &  2.985 (2.976)  &  0.960 (0.956)  &  1.31e-01 (1.35e-01) \\
 & 8  &  64  &   10.022 (10.022)  &  3.031 (3.025) &   0.962 (0.960)  &  1.05e-01 (1.08e-01) \\
 & 8  &  72  &   10.009 (10.009) &   3.035 (3.029) &   0.963 (0.961) &   1.02e-01 (1.06e-01) \\
 & 8  &  80   &  10.015 (9.998)  &  3.046 (0.731)  &  0.962 (0.070)  &  9.63e-02 (1.50e\;00) \\
& 12  &  48  &   10.046 (10.046)  &  3.200 (3.195)  &  0.941 (0.939) &   3.02e-02 (3.04e-02) \\
& 12  &  96  &   10.041 (10.365)  &  3.212 (3.578)  &  0.940 (0.800)  &  2.69e-02 (2.07e-01) \\
& 12  & 108  &   10.038 (10.755)  &  3.214 (8.397)  &  0.938 (3.062)  &  2.52e-02 (3.24e\;00) \\
& 12  & 120  &   10.037 (15.365)  &  3.217 (17.140)  &  0.937 (8.341) &   2.48e-02 (9.31e\;00) \\
	 &	&  &  &   &    & \\
	\hline
\multicolumn{3}{c}{Prices in \cite{Hout15}} & 10.004 & 3.213 & 0.931 &  \\
\multicolumn{3}{c}{Prices in \cite{oosterlee11}}  & 9.996 & 3.208 & 0.928 &  \\
\noalign{\smallskip}\hline
	\end{tabular}
	}
\end{table}

\section{Conclusions}

In this paper, we have applied the symmetric interior penalty dGFEM for solving the European and American option prices under Heston model. We have shown
the non-smooth boundary and initial conditions for various option pricing models can be handled in a natural way by the dGFEM in combination with the Rannacher smoothing in time for various European option pricing models. The adaptive grid based on a posteriori error estimate demonstrates the performance of the dGFEM for convection dominated Heston model. Due to the convective term in the Heston model, the stiffness matrix resulting from dGFEM discretization is nonsymmetric and nonnormal. The nonnormality of this matrix affects negatively the convergence of iterative methods like the PSOR. The left-right norm preconditioner transforms the stiffness matrix resulting from dGFEM discretization into a normal matrix, which accelerate the convergence of PSOR. The numerical results for European and American options using different parameters agree well with those in the literature.

\section*{Acknowledgments}

The authors would like to thank Yeliz Yolcu Okur for the comments and suggestions that helped to improve the manuscript and  for constructive comments of the anonymous referees which helped to improve the paper.


\end{document}